\documentclass[aps,a4paper]{revtex4}
\usepackage[colorlinks=true, pdfstartview=FitV, linkcolor=blue, citecolor=red, urlcolor=magenta,breaklinks=true]{hyperref}
\usepackage[T1,T5]{fontenc}
\usepackage{graphicx}
\usepackage[all]{xy}
\usepackage{amsmath}
\usepackage{amssymb}
\usepackage{color}
\usepackage{subeqnarray}
\usepackage{subfigure}
\usepackage{epstopdf}

\newcommand{\bb}{\bibitem}
\newcommand{\bes}{\begin{subequations}}
\newcommand{\ees}{\end{subequations}}
\def\ben{\begin{eqnarray}}
\def\een{\end{eqnarray}}
\newcommand{\bens}{\begin{subeqnarray}}
\newcommand{\eens}{\end{subeqnarray}}

\def\be{\begin{equation}}
\def\ee{\end{equation}}

\def\tanh{\text{tanh}}
\def\sech{\text{sech}}

\def\cos{\text{cos}}
\def\arccot{\text{arccot}}

\begin{document}
\title{Kinklike structures in an arcsin real scalar dynamics} 
\author{Diego R. Granado$^{a,b}$}
\author{Elisama E.M. Lima$^{c}$}
  
\affiliation{$^{a}$ Institute of Research and Development, Duy T{a}n University, Da Nang 550000, Viet Nam}
\affiliation{$^{b}$ Faculty of Natural Sciences, Duy T{a}n University, Da Nang 550000, Viet Nam}
\affiliation{$^{c}$ Federal Institute of Education, Science and Technology of Bahia, Barreiras-BA, 47808-006, Brazil}


\begin{abstract}
In this paper, we analyze kink-like analytical solutions in a real scalar theory with an arcsin dynamics inspired by the arcsin electrodynamics presented in \cite{Kruglov:2014iwa}. This analysis is done by means of the first-order formalism. This formalism provides a framework where the equations of motion can be simplified by preserving the linear stability of the theory. In this work, the 
deformation procedure is implemented with the aim of finding exact solutions in systems with generalized dynamics. Along the paper, we explore how the first-order formalism is implemented in the arcsin kinetics and how such a term influences the kink-like solutions. As a part of the result of our paper, we show that the kink-like solutions are similar to the ones obtained in the standard scalar kinetic theory. We also show that the extra parameter, that controls the non-linearities of the model, plays an essential role in the energy densities and stability potentials. These quantities vary according to this parameter. The goals here are to show how the first-order framework is implemented in this arcsin scenario and to present the analytical kink-like solutions that can be found by means of the first-order framework and deformation method.

\end{abstract}
\maketitle


\section{Introduction}
\label{intro}
In order to solve the electron self-energy problem, Born and Infeld (BI) proposed a non-linear electrodynamics  model \cite{Born:1934gh}. As part of the model, an extra mass dimensional parameter was introduced to deal with the energy divergence problem. This extra parameter introduces a cutoff to the theory. The standard (linear) dynamics can be recovered for small/large values of this parameter,  depending on how it was implemented in the Lagrangian. This idea presents the picture that non-linear electrodynamics models can be a solution to deal with the electron self-energy problem. Thus, along the years, following the same idea presented by BI, other examples of non-linear electrodynamics were considered \cite{Dirac:1962iy,Kruglov:2007bh,Kruglov:2001dp,Kruglov:2007zr,Gaete:2013dta,Kruglov:2014hpa,Kruglov:2014iwa}. 

Topological defects are structures of great interest in the whole physics community \cite{tops}. The simplest one is a topological structure known as \textit{kink}. It arises in one-dimensional static scalar classical field theory as solutions of the equations of motion. 
In order to analytically study the presence of kink-like structures, in a real scalar model inspired by the non-linear electrodynamics proposed in Ref.~\cite{Kruglov:2014iwa}, we make use of the first-order formalism \cite{bogomol,FOFGD}. The first-order formalism was applied for a great variety of systems \cite{FIRST}. Concerning to non-linear dynamics scalar models inspired by non-linear electrodynamics models, in the Ref.~\cite{Bazeia:2017mnc}, the authors worked out the first-order formalism to find kink-like structures in the DBI scalar model. As proposed in \cite{Bazeia:2017mnc}, the idea here is to use the deformation procedure \cite{Bazeia:2002xg}. This procedure is useful to look for exact solutions in generalized scalar models \cite{BLM2014}.  

The goal of this paper is, by means of the path introduced in Ref. \cite{Bazeia:2017mnc}, to find kink-like structures in scalar models inspired by the non-linear electrodynamics proposed by the Ref. \cite{Kruglov:2014iwa}. Along this paper, we deal with two categories of scalar field models: one described
by modified polynomial interactions and another one related to non-polynomial interactions. The dynamics and potential terms are such that the non-linearities preserve the kink-like behavior, but modify the energy density and stability potential.

This paper is organized as follows: in the section \ref{sec-1}, we review the first-order formalism for a single real scalar in a general framework; in the section \ref{sec-2}, we implement the first-order formalism for the arcsin scalar theory and we describe a model supporting topological solutions with non-standard dynamics; in the section \ref{defp}, we incorporate the deformation procedure to find analytical solutions for several models described by the arcsin dynamics; in the section \ref{sec-com}, we conclude and comment our results.

\section{First-Order Formalism} \label{sec-1}

In this section, we present a short review of the first-order formalism \cite{FOFGD}, for the case of a single real scalar field in a non-standard dynamics. The most general action for that situation in $(1,1)$ space-time dimensions reads
\be
\label{lagran0}
S=\int d^2x {\cal L}\left(\phi,X\right),
\ee
where $X$ reads
\be
X=\frac{1}{2}\partial_\mu\phi\partial^\mu\phi.
\ee
The equation of motion obtained from the action \eqref{lagran0} reads:
\be
\partial_\mu({\cal{L}}_{X}\partial^\mu\phi)={\cal{L}}_{\phi},
\label{eom0}
\ee

The energy-momentum tensor reads
\be
T_{\mu\nu}={\cal{L}}_{X}\partial_\mu\phi\partial_\nu\phi-g_{\mu\nu}{\cal{L}}.
\ee
Since we are working with static field configurations; we can write the energy-momentum components in the form:
\begin{eqnarray}
\rho(x)&=&-{\cal{L}}\\
\tau(x)&=&{\cal{L}}_{X}\phi'^2+{\cal{L}}.
\label{stress}
\end{eqnarray}
The equation of motion becomes
\be
\left({\cal{L}}_{X}+2{\cal{L}}_{XX}X\right)\phi''=2X{\cal{L}}_{X\phi}-{\cal{L}}_{\phi}.
\label{staticequationofmotion}
\ee
This equation can be integrated to give:
\be
{\cal{L}}-2{\cal{L}}_{X}X=C,
\ee
which can be identified as the spatial component of the energy-momentum tensor \eqref{stress}, where $C$ is an integration constant. The stressless condition demanded from stability gives
\be
{\cal{L}}-2{\cal{L}}_{X}X=0.
\label{fo}
\ee
From this constraint the energy density can be written as
\be
\rho(x)=-{\cal{L}}={\cal{L}}_{X}\phi'^2.
\label{rho}
\ee
Furthermore, the formerly equation of motion can be written as $({\cal{L}}_{X}\phi')'=-{\cal{L}}_{\phi}$; and we can define a function $W=W(\phi)$ satisfying this second order equation, as
\be
{\cal{L}}_{X}\phi'=W_{\phi}.
\label{w}
\ee
Consequently, the function $W$ must obey
\be
W_{{\phi}{\phi}}\phi'=-{\cal{L}}_{\phi},
\label{here}
\ee
and the first-order equation \eqref{fo} leads to 
\be \label{gpot}
{\cal{L}}+\phi'W_{\phi}=0.
\ee
The energy density can be written as
\be
\rho(x)=W_{\phi}\phi'=\frac{d W}{dx},
\label{rhol}
\ee
which gives the total energy 
\begin{eqnarray}
E=\Delta W&=&W(\phi(\infty))-W(\phi(-\infty)).
\end{eqnarray}
Until this point, we have presented the framework as much general as it is possible, by considering only the most general action as in action \eqref{lagran0}. In \cite{Granado:2019bky}, the author show that the first-order formalism can be implemented in the real scalar field setup in the most general case possible. As in the section \ref{sec-2}, we analyze kink-like structures in a specific model, from this point we chose the Lagrangian density to be of the following type
\be
{\cal L}\left(\phi,X\right)=F(X)-V(\phi).
\label{linearlagren}
\ee
where the implementation of the first-order formalism for the Lagrangian \eqref{linearlagren} can be found in \cite{Bazeia:2017irs}. For $F(X)=X$, the standard dynamics is recover. Now, the equation of motion \eqref{staticequationofmotion} reduces to the form
\be
(2XF_{XX}+F_{X})\phi''=V_\phi.
\label{1}
\ee
The definition of $W(\phi)$ obtained from Eq.~\eqref{w} reads
\be
\label{2}
F_X\phi'=W_\phi,
\ee
and the equation \eqref{here} become:
\be
\label{3}
V_\phi=W_{\phi\phi}\phi'.
\ee
From equation \eqref{rho}, we get the energy density
\be
\rho(x)=F_{X}\phi'^2=W_\phi \phi'.
\label{rhoFx}
\ee
In order to verify the linear stability of the model, we make use of the perturbation theory around the static solution: $\phi(x,t)=\phi_s(x)+\eta(x,t)$. In this situation, 
\ben
X&=&-\frac12\phi_s'^2-\phi_s'\eta' \\
&=&X_s+\overline{X}
\een
with $\overline{X}=-\phi_s'\eta'$, and we are considering terms up to first order in $\eta$. We expand the functions bellow
\ben
V_\phi&=&V_{\phi_s}+V_{\phi_s\phi_s}\eta, \\
F_X&=&F_{X_s}+F_{X_s X_s}\overline{X};
\een
and from the equation of motion \eqref{eom0}, we have
\be
F_{X_s}\ddot{\eta}-\partial_x \left(F_{X_s}\eta'+F_{X_s X_s}\, \overline{X}\, \phi' \right)+V_{\phi_s\phi_s}\eta =0.
\ee
Supposing the fluctuation of the kind $\eta(x,t)=\eta(x)\cos(\omega t)$, one gets 
\be
\label{eqpert}
-\left[\left(F_{X_s}+2X_s F_{X_s X_s}\right)\eta'\right]'= \left( F_{X_s}\omega^2-V_{\phi_s\phi_s}\right)\eta,
\ee
which is the Sturm-Liouville equation of the form: $-\left[a(x)\eta'\right]'=b(x)\eta$. In order to get a Schr\"odinger-like equation, one can define the new set of variables 
\begin{eqnarray}
\label{4}
dx=Adz \qquad \mbox{and}\qquad
\eta=\frac{u}{\sqrt{F_XA}},
\label{eta}
\end{eqnarray}
where
\be
A^2=\frac{2X_sF_{X_sX_s}+F_{X_s}}{F_{X_s}}.
\label{A}
\ee
As follows, the perturbed equation \eqref{eqpert}  can be reassembled in the form: $Hu(z)=\omega^2u(z)$, such that
\be
\left(-\frac{d^2}{dz^2}+U(z)\right)u(z)=\omega^2u,
\label{schrodingerlikequation}
\ee
with the stability potential:
\be
U(z)=\frac{(\sqrt{F_{X_s}A})_{zz}}{\sqrt{F_{X_s}A}}+\frac{V_{\phi_s\phi_s}}{F_{X_s}}.
\label{stabilitypotential}
\ee
From the equation \eqref{schrodingerlikequation}, we have the zero mode equation:
\be
\left(-\frac{d^2}{dz^2}+U(z)\right)u_0(z)=0.
\ee
Then, the stability potential can be written as:
\be
U(z)=\frac{1}{u_0(z)}\frac{d^2}{dz^2}u_0(z).
\ee
Furthermore, comparing the Sturm-Liouville equation \eqref{eqpert} for $\omega=0$ with the derivative of the static equation of motion \eqref{1}, we find out the zero mode: $\eta_0=\phi'$. Using this fact and the Eq.~\eqref{eta}, we can rewrite the stability potential as:
\be
U(z)=\left. \left[\frac{ A(x)}{\phi'\sqrt{F_{X_s} A}}\frac{d}{dx}\left(A(x) \frac{d(\phi'\sqrt{F_{X_s}A})}{dx}\right)\right] \right|_{x=x(z)}.
\label{u(z)}
\ee

\section{ARCSIN Model} \label{sec-2}

By means of the formalism presented in the previous section, in this section, we introduce the first-order formalism for a scalar model inspired by the non-linear electrodynamics presented in the Ref.~\cite{Kruglov:2014iwa}. In this scenario, the Lagrangian reads:
\be \label{lagranarcsin}
\mathcal{L}=\frac{1}{2\,\beta}\arcsin(2\,\beta X)-V(\phi),
\ee
where the parameter $\beta$ is real and positive. Particularly, for small $\beta$ the kinetic term reads:
\be
\frac{1}{2\,\beta}\arcsin(2\,\beta X)=X+\frac23\,\beta^2 X^3+\dots,
\label{expansion}
\ee
which returns to the standard model, when $\beta=0$. The Lagrangian \eqref{lagranarcsin} furnishes the following static equation of motion 
\be
\frac{d}{dx}\left(\frac{\phi'}{\sqrt{1-({\beta} \phi'^2)^2}}\right)=V_\phi.
\ee
As in \eqref{w}, the function $W$ will be defined as
\be
W_\phi=\frac{\phi'}{\sqrt{1-({\beta} \phi'^2)^2}},
\label{wphi}
\ee
whose solution for $\phi'^2$ is
\be
\phi'^2=\frac{1}{2{\beta}^2W_\phi^2}\left(\sqrt{1+4{\beta}^2W_\phi^4}-1\right),
\label{solutionphip}
\ee
Following the steps to reach \eqref{rhoFx}, we have that
\be
\label{rhoarcsin}
\rho(x)=\frac{\phi'^2}{\sqrt{1-({\beta} \phi'^2)^2}}.
\ee
Using the equation \eqref{gpot}, we obtain the following form for the potential $V(\phi)$ 
\be \label{pote}
V(\phi)=-\frac{1}{2\,\beta}\arcsin{(\beta\phi'^2)}+W_{\phi}\phi'.
\ee
This  potential in terms of  $W_\phi$ can be obtained using the Eq.~\eqref{solutionphip}, so
\ben
V(\phi)&=&-\frac{1}{2\,\beta}\arcsin{\left(\frac{1}{2{\beta}W_\phi^2}\left(\sqrt{1+4{\beta}^2W_\phi^4}-1\right)\right)} \nonumber \\
&& + \frac{1}{\sqrt{2}\beta}\left(\sqrt{1+4{\beta}^2W_\phi^4}-1\right)^{1/2}.
\een
In the case of standard dynamics, the states satisfying the first-order formalism are obtained by a BPS compatible potential: $V=\frac{1}{2}W_\phi^2$ \cite{bogomol}. 
In the context presented here, the standard dynamics results are recovered for small values of $\beta$, as it was pointed out by the expansion \eqref{expansion}. In this way, the solution \eqref{solutionphip}, expanded for small $\beta$ recovers the standard scenario: $\phi'=\pm W_\phi$ \cite{FOFGD}.

\subsection{Modified $\phi^4$}

\begin{figure}%
\centering
\includegraphics[scale=0.3]{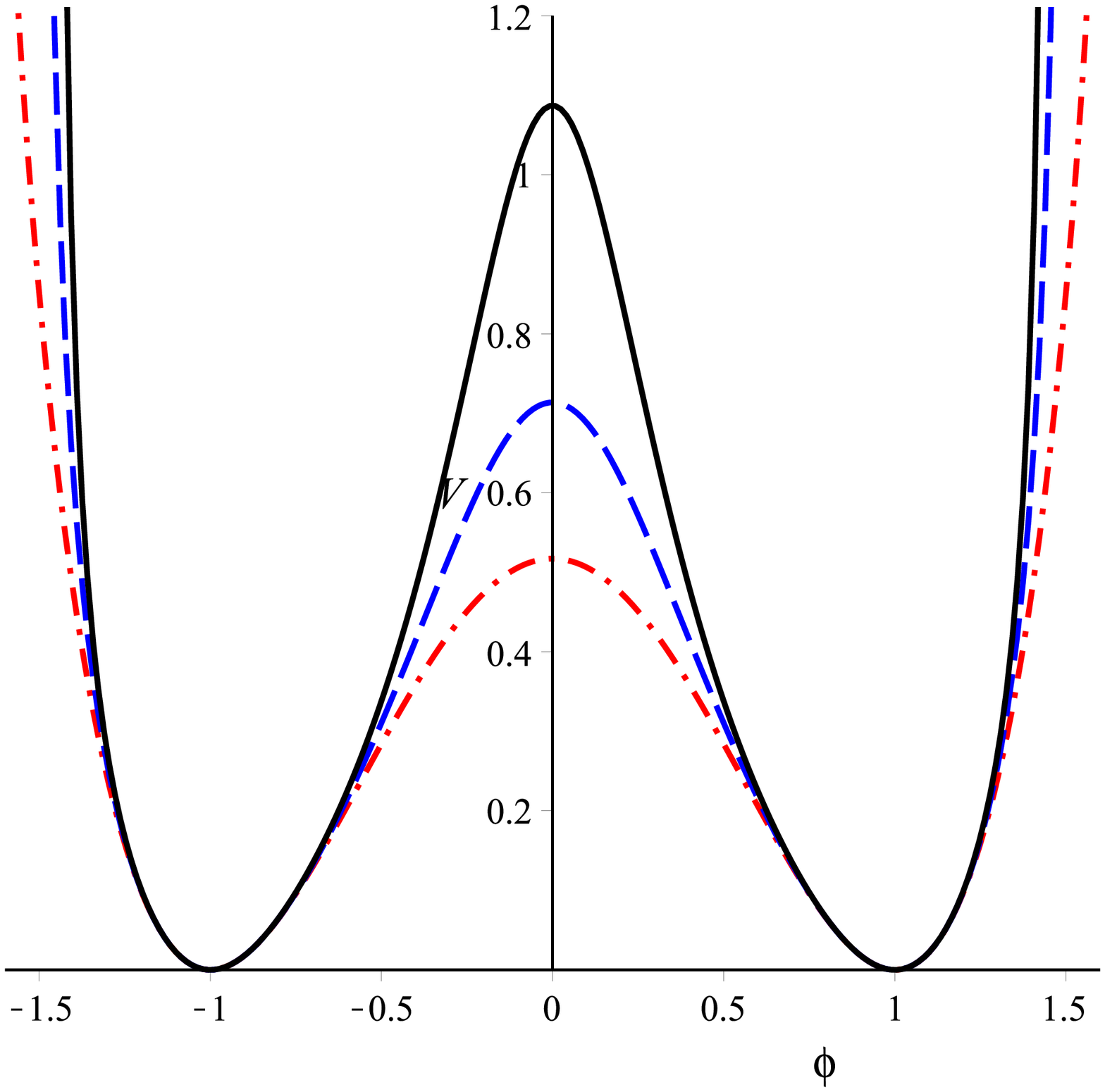}
\includegraphics[scale=0.3]{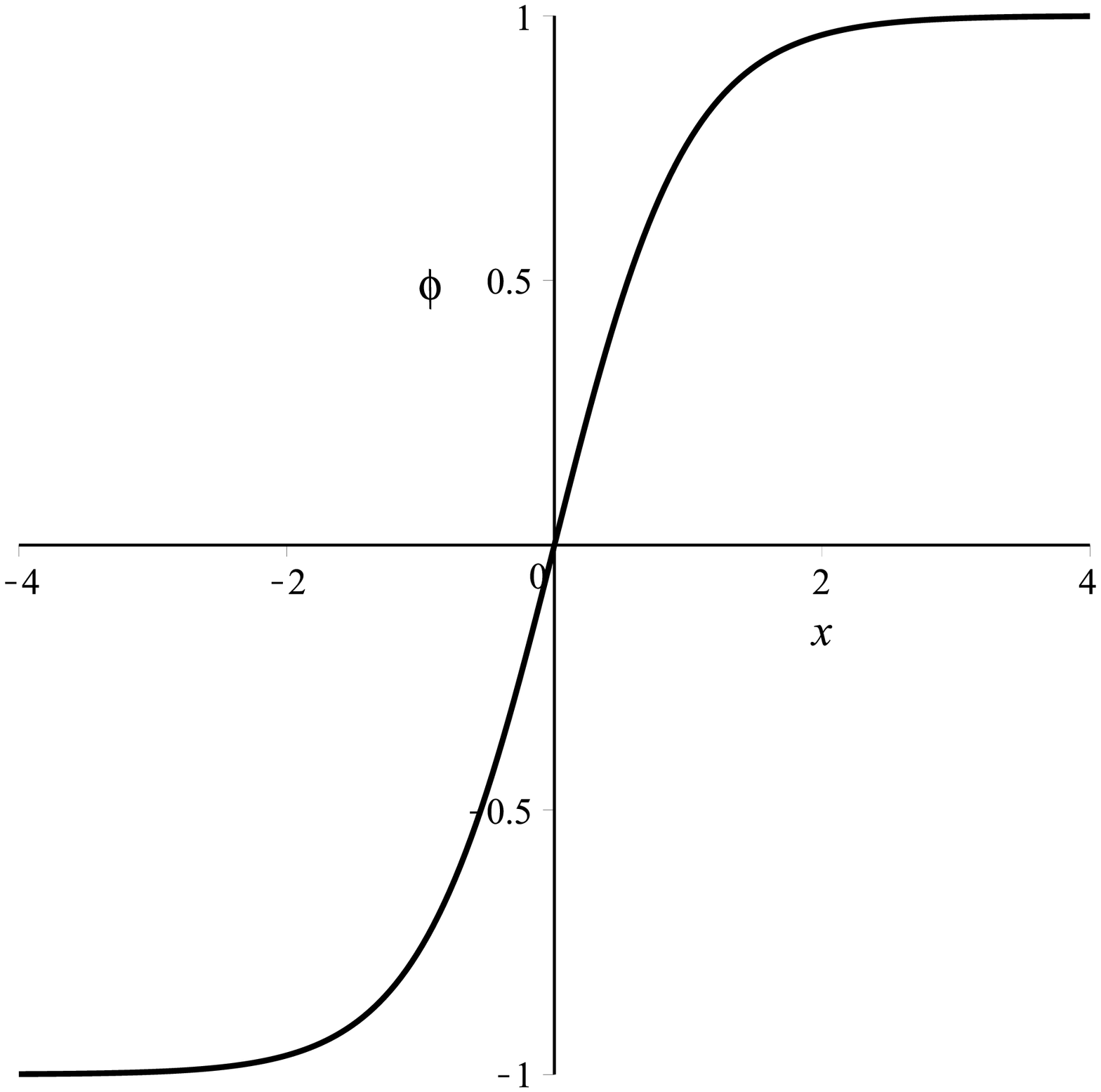}
\includegraphics[scale=0.3]{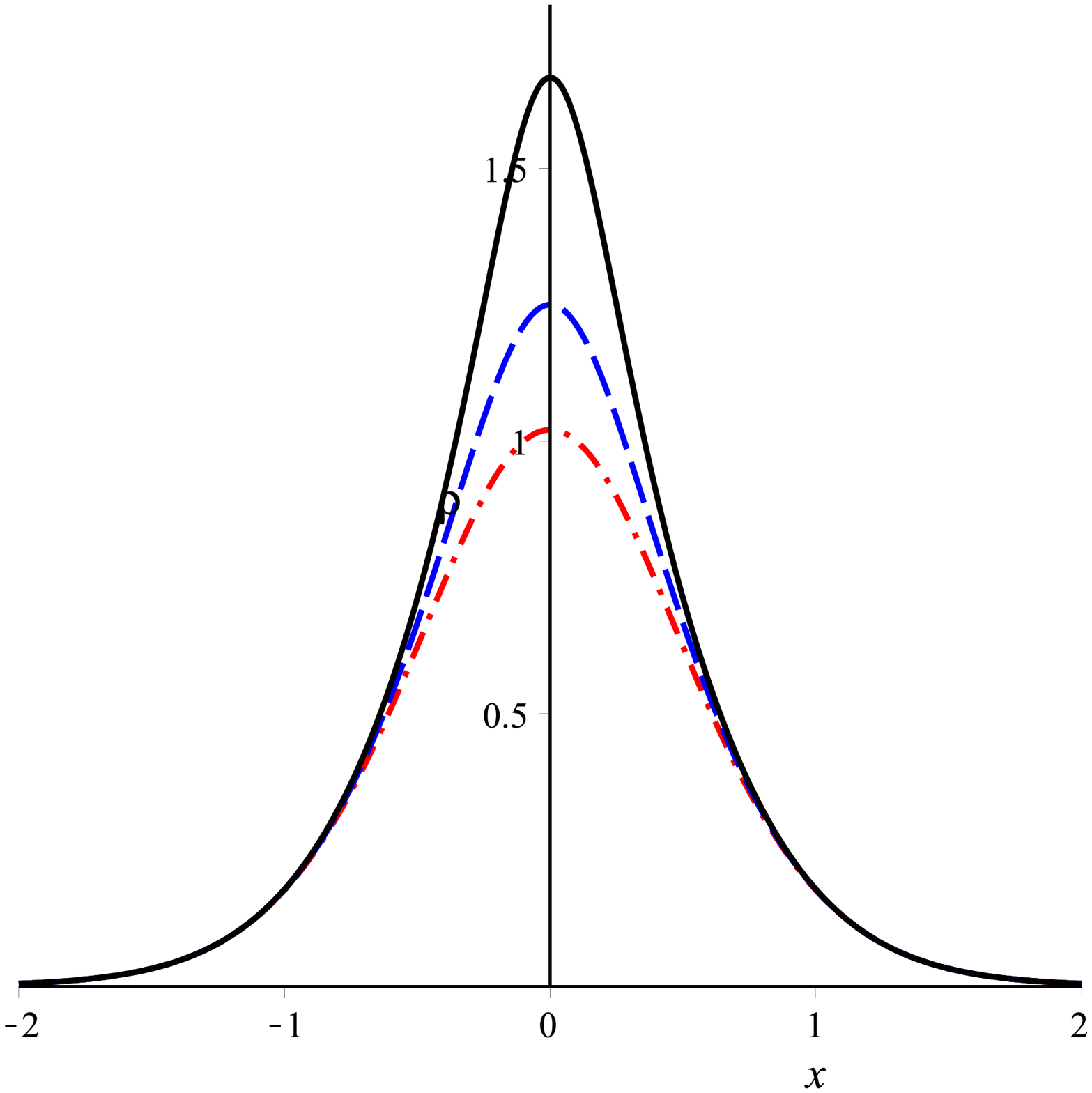}
\includegraphics[scale=0.3]{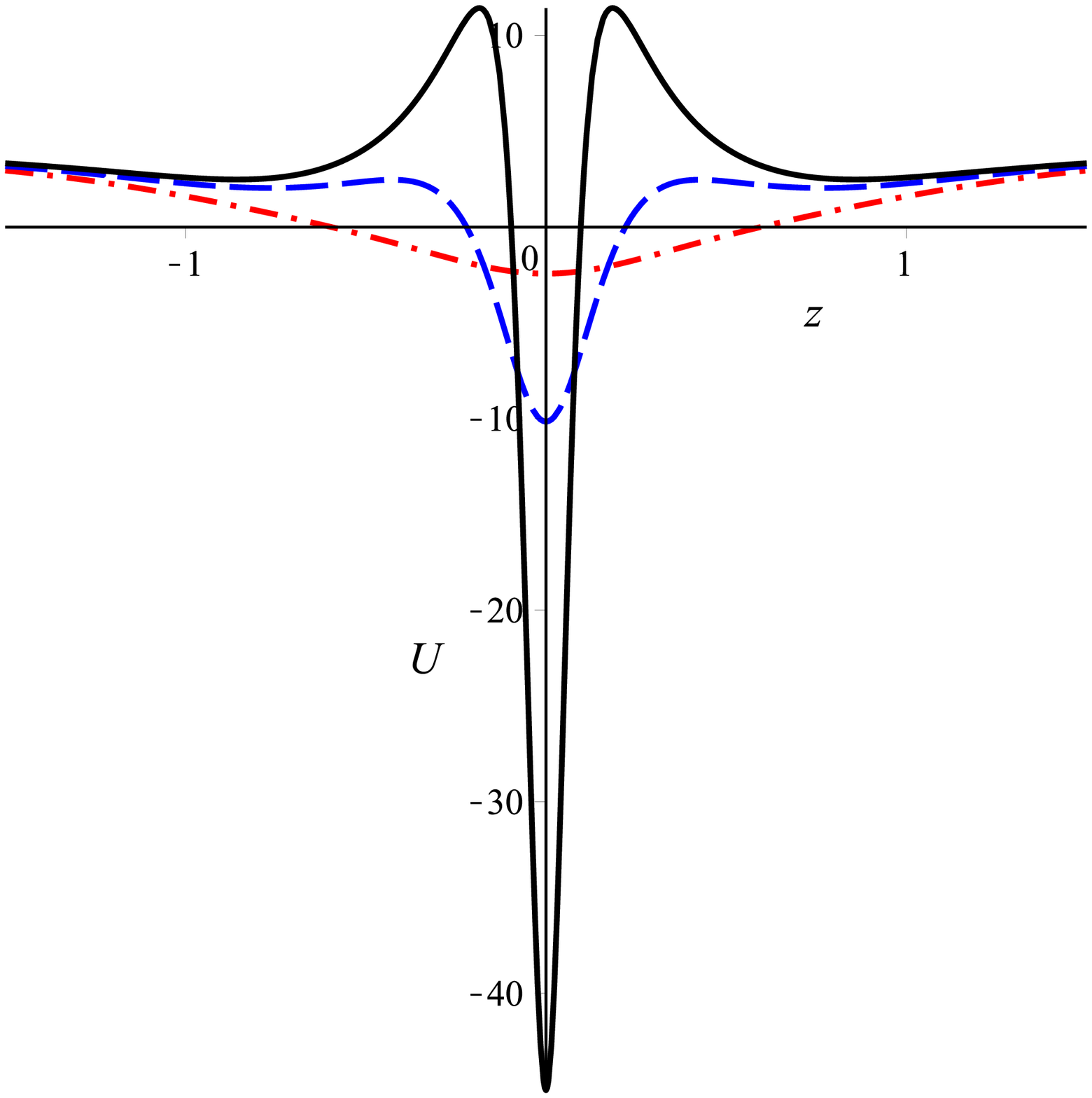}
\caption{In the figures above we have: the potential \eqref{potp4}, the kink solution \eqref{phikink}; the energy density \eqref{energydensity} and the stability potential $U(z)$. For ${\beta} = 0.2, 0.6, 0.8$, represented by dashdotted (red),  dashed (blue), and solid (black)  lines,  respectively.}
\label{plot}
\end{figure}

As mentioned at the introduction, kinks are the most common topological structures that can be found in a one-dimensional static scalar theory. Here, we show how the non-standard arcsin dynamics can influence the energy density, potential, and stability potential of these topological structures.  

In order to find kink solutions, we begin with the following proposition:
\be
W_\phi=\pm \frac{1-\phi^2}{\sqrt{1-{\beta}^2(1-\phi^2)^4}},
\label{wkink}
\ee
and we can verify that the standard results are recovered for $\beta<<1$, where $W_\phi =\pm(1-\phi^2)$. Besides that, from the equation \eqref{solutionphip}, we have $\phi'=\pm(1-\phi^2)$; so the scalar field potential \eqref{pote} in terms of $\phi$ is
\be
V(\phi)=-\frac{1}{2\beta}\arcsin\left(\beta(1-\phi^2)^2\right)+\frac{(1-\phi^2)^2}{\sqrt{1-\beta^2(1-\phi^2)^4}}.
\label{potp4}
\ee
The behavior of the potential \eqref{potp4} for different values of $\beta$  is presented in the first plot of the Fig.~\ref{plot}; there $V(\phi)$ varies slightly for small values of $\beta$,  and  meaningful changes appear at the potential maximum as $\beta \rightarrow 1$. In fact, at the potential maximum $\phi=0$, we have
\be
V(0)=-\frac{1}{2\beta}\arcsin(\beta)+\frac{1}{\sqrt{1-\beta^2}},
\ee
that indicates a range for the theory parameter, $0<\beta<1$, to obtain real results for the potential at $\phi=0$. Moreover, when $\beta<<1$, the potential becomes the usual $\phi^4$-model in a standard perspective, \textit{i.e.}, $V(\phi) \approx \frac{1}{2}(1-\phi^2)^2$ \cite{tops}. 

The kink solution $\phi(x)$ reads:
\be
\phi(x)=\tanh(x);
\label{phikink}
\ee
which is independent from the parameter $\beta$. There is also a similar solution towards the opposite direction, called antikink, obtained by the transformation: $x \rightarrow - x$.  The kink solution $\phi(x)$ transits between $-1$ and $1$, which are the minima of the potential $V(\phi)$, constituting the topological sector of the theory, as can be seen in the first and second plot of Fig.~\ref{plot}.

Concerning the density energy $\rho(x)$ and the stability potential $U(z)$, by means of the equation \eqref{rhoarcsin}, the density energy reads:
\be
\rho(x)=\frac{\sech(x)^4}{\sqrt{1-{\beta}^2\sech(x)^8}}.
\label{energydensity}
\ee
This expression shows us that there is a singularity on $x=0$ for $\beta=1$; thus, as seen previously, it is required that $\beta<1$ for meaningful results. For $\beta<<1$, we have $\rho(x)\approx\sech(x)^4$ which is the standard $\phi^4$ energy density, as expected by the expansion \eqref{expansion}. From the third plot on Fig.~\ref{plot}, we can see $\rho(x)$ changing as the $\beta$ value varies. 
 
By means of the expression \eqref{u(z)}, it can be seen that a numerical integration is required to find $z(x)$,  as the change of variables \eqref{eta} carries a non-analytical integral equation,
\be
z(x)=\int{\frac{dx}{A(x)}},
\ee
where
\be
A(x)=\left(\frac{1+\beta^2\sech(x)^8}{1-\beta^2\sech(x)^8}\right)^{1/2}.
\ee
As a result, we have the profile $U(z)$, presented on the last plot of the Fig.~\ref{plot}. As the parameter $\beta$ approaches to unity, the stability potential presents a volcano-like behavior, originated from the non-linearities inherent of our dynamics; volcano behaviors can also be found in models with a warped geometry \cite{brane,Barbosa}. Furthermore, the point $x=0$, for $\beta=1$ is singular, due to the following quantity on the equation \eqref{u(z)}
\be
\sqrt{F_XA}=\left(\frac{\sqrt{1+\beta^2\sech(x)^8}}{1-\beta^2\sech(x)^8}\right)^{1/2},
\ee
and this reinforces the constraint $\beta<1$. For $\beta<<1$, the potential stability reads:
\be
U(x)\approx 4-6\,\sech^2(x),
\ee
where two bound states can be found: one zero mode and one excited state \cite{teller}.

\section{Deformation Procedure}
\label{defp}

The deformation method for modified dynamics was constructed in Ref.~\cite{BLM2014}. This method provides analytical solutions for new models related by a deforming function $f(\chi)$, which connects a system of a potential $V(\phi)$, supporting analytical solutions, with another $\tilde V(\chi)$, which is the the desired one where the solutions can be found. 

Accordingly, we introduce a new system satisfying the first-order construction presented in Sec.~\ref{sec-2}. The deformed Lagrangian $\mathcal{L}=\mathcal{L}(\chi,Y)$ describing the scalar field $\chi(x, t)$ reads
\be
\mathcal{L}=\frac{1}{2\,\beta}\arcsin(2\,\beta Y)-\tilde{V}(\chi),
\ee
with $Y=\frac12\partial_\mu\chi\partial^\mu\chi$. The novel potential is
\be
\tilde{V}(\chi)=-\frac{1}{2\,\beta}\arcsin{(\beta\chi'^2)}+W_{\chi}\chi',
\ee
where
\be
W_\chi=\frac{\chi'}{\sqrt{1-({\beta} \chi'^2)^2}};
\label{wchi}
\ee
and the first-order equation reads
\be
\chi'^2=\frac{1}{2{\beta}^2W_\chi^2}\left(\sqrt{1+4{\beta}^2W_\chi^4}-1\right),
\label{solutionchi}
\ee
From the equation above, we can write $\chi'$ in terms of an unknown function $S(\chi)$, such as:  $\chi'=S(\chi)$; and from  Eq.~\eqref{solutionphip} we express: $\phi'=R(\phi)$. Relating those models through a differentiable deforming function $f(\chi)$, in such way that $\phi \rightarrow f(\chi)$, we find 
\be
S(\chi)=\frac{R(\phi\rightarrow f(\chi))}{f_\chi}.
\ee  
The deformed potential can be written as $ \tilde{V}(S(\chi))$, then
\be
\tilde{V}(\chi)=-\frac{1}{2\,\beta}\arcsin{(\beta S^2)}+\frac{S^2}{\sqrt{1-(\beta S^2)^2}}.
\label{deformmodel}
\ee

Furthermore, using the model \eqref{potp4}, which has known static solution $\phi(x)= \pm \tanh(x)$, we have
\be
S(\chi)=\frac{1- f^2}{f_\chi},
\label{schi}
\ee 
and the solution, for the deformed model \eqref{deformmodel}, is given by the inverse deforming function: $\chi(x)=f^{-1}(\phi(x))$. This allows us to find and solve new generalized models described by real scalar fields.

The energy density in terms of $S(\chi(x))$ is obtained by
\be
\rho(x)=\frac{S^2}{\sqrt{1-\beta^2 S^4}}.
\label{rhodef}
\ee
In the next examples, we introduce some known deforming functions to analyze effects of the arcsin dynamics in novel field models supporting topological structures.  

\subsection{Modified $\phi^6$}

\begin{figure}%
\centering
\includegraphics[scale=0.3]{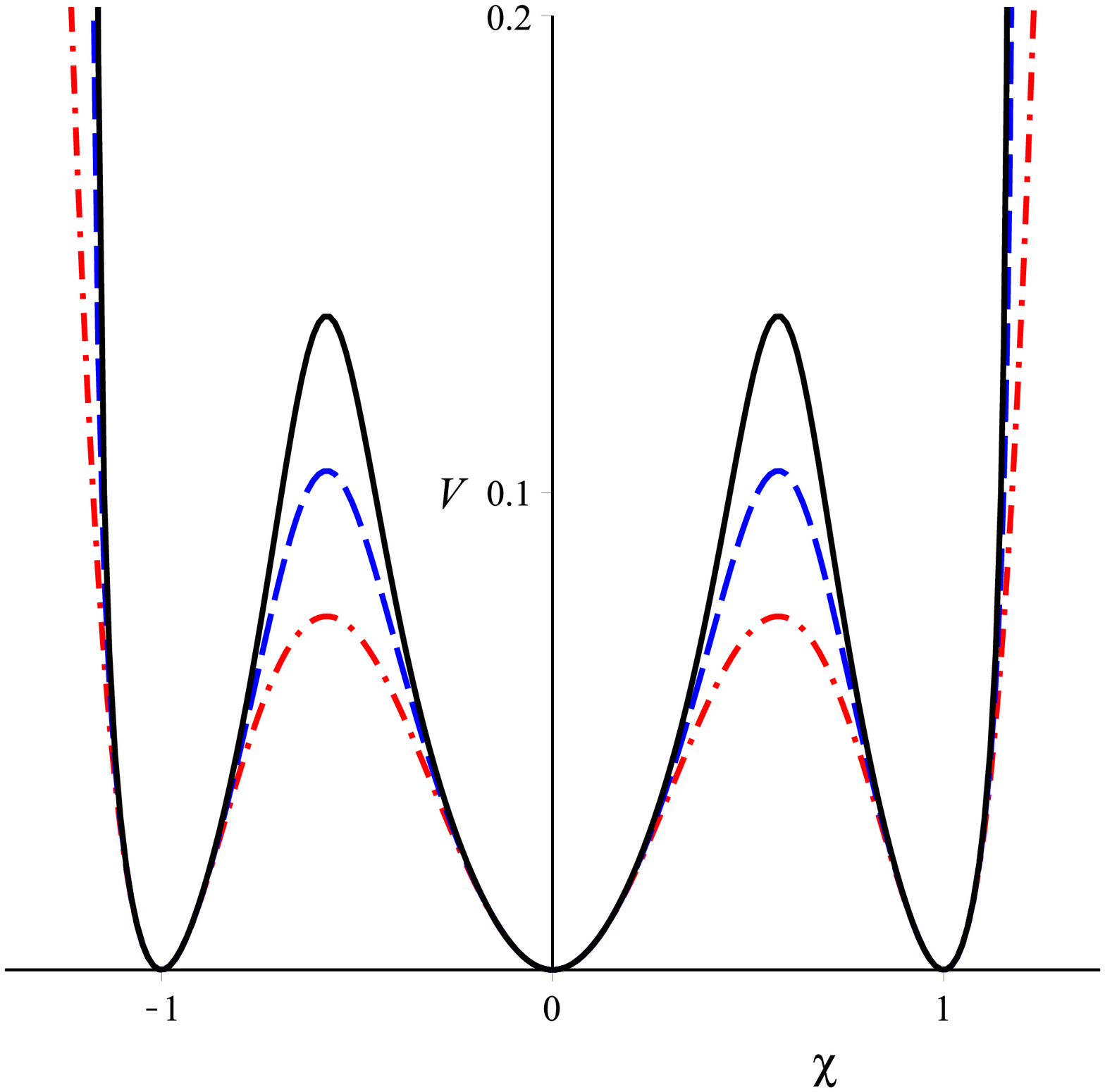}
\includegraphics[scale=0.3]{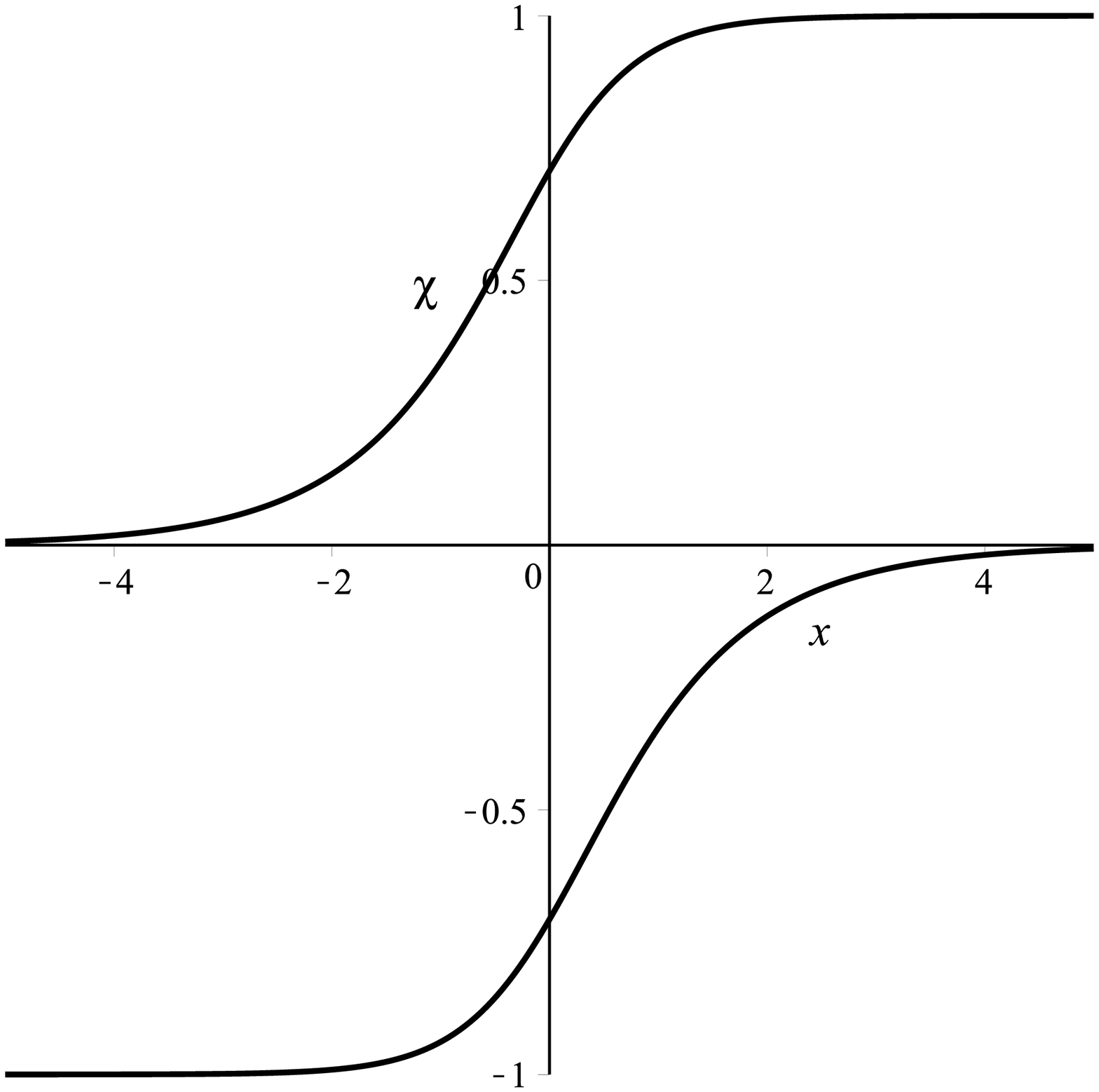}
\includegraphics[scale=0.3]{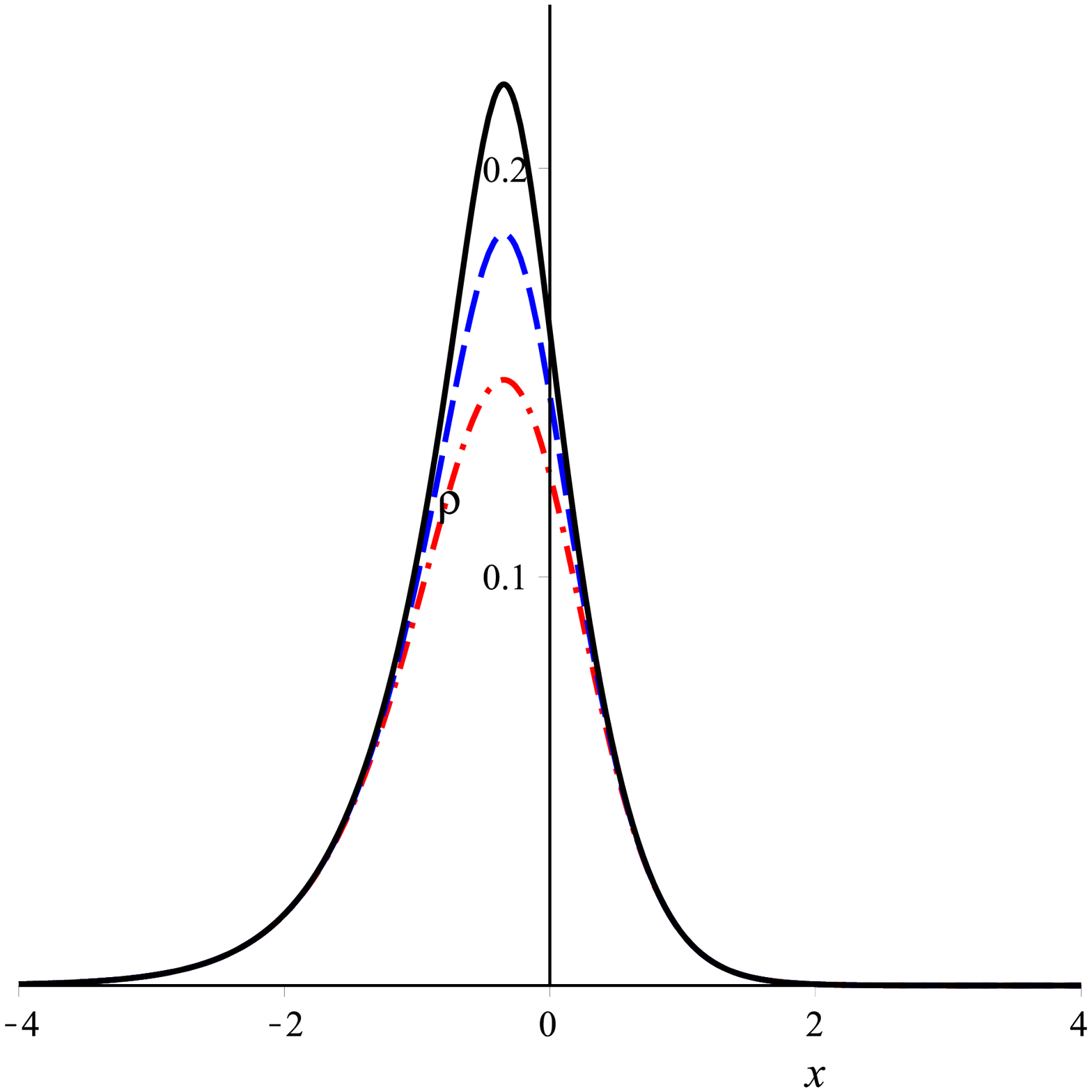}
\includegraphics[scale=0.3]{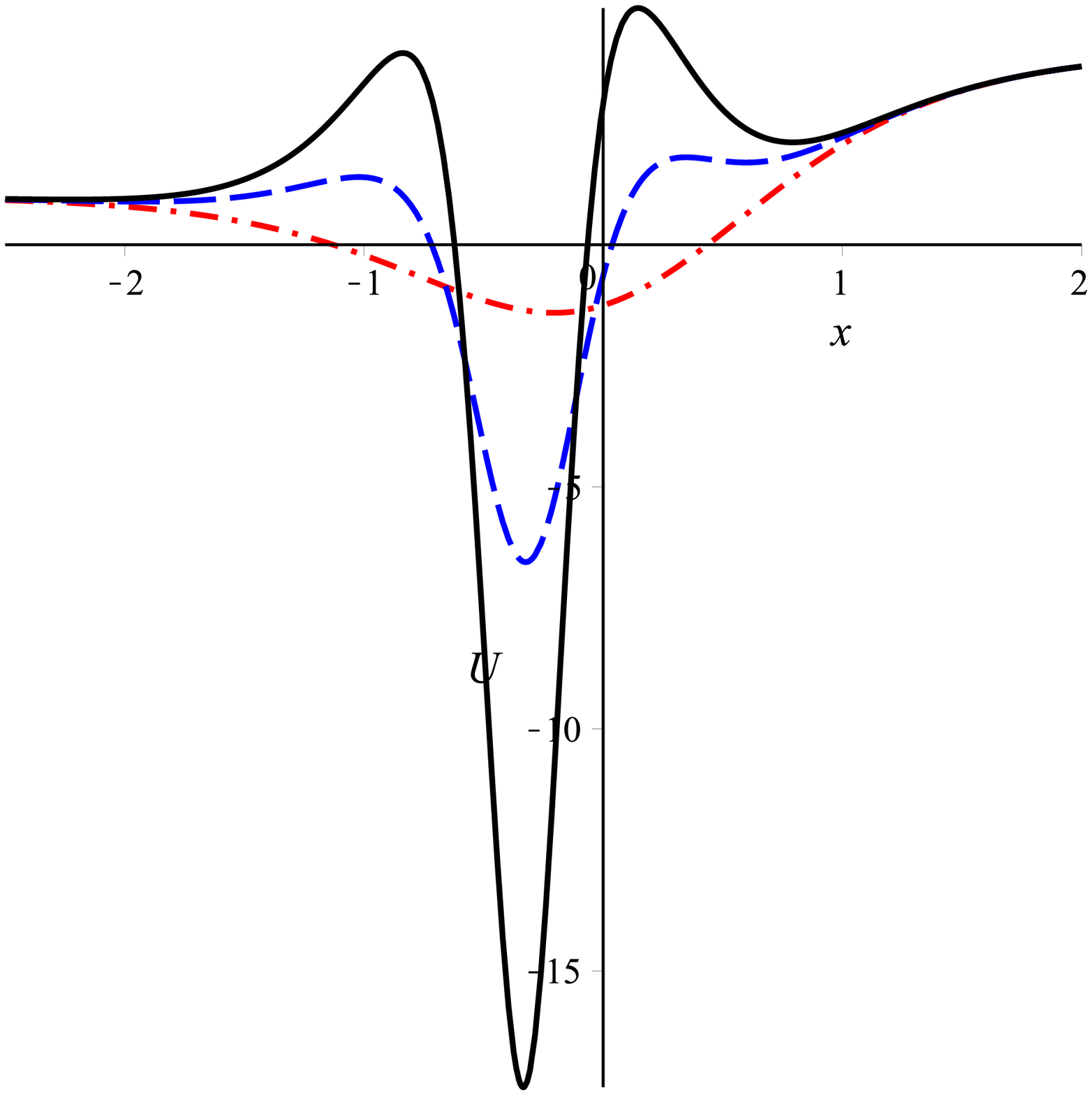}
\caption{The potential \eqref{potd1}, the kink solutions \eqref{sold1}, the energy density \eqref{densd1}  and the stability potential \eqref{u(z)} for the positive kink solution. For ${\beta} = 0.2, 3, 5$, represented by dash-dotted (red), dashed (blue), and solid (black)  lines, respectively.}
\label{plotp6}
\end{figure}

Using the following deforming function: $f(\chi)=2\chi^2-1$, we have
\be
S(\chi)=\chi(1-\chi^2),
\ee
and the deformed potential is
\be
\label{potd1}
\tilde{V}=-\frac{1}{2\,\beta}\arcsin{(\beta \chi^2(1-\chi^2)^2)}+\frac{ \chi^2(1-\chi^2)^2}{\sqrt{1-\beta^2\chi^4(1-\chi^2)^4}},
\ee
where \eqref{potd1} is a $\phi^6$-like model with minima at $\chi_{min}=0,\pm 1$ and maxima at $\chi_{max}=\pm 1/\sqrt{3}$, as it can be seen on the first plot of Fig.~\ref{plotp6}. From the maxima points, we have
\be
\tilde{V}(\chi_{max})=-\frac{1}{2\beta}\arcsin\left(\frac{4 \beta}{27}\right)+\frac{4}{\sqrt{729-16\beta^2}},
\ee
which means that $0<\beta<27/4$. 

The solutions obtained by $\chi(x)=f^{-1}(\phi(x))$ are 
\be
\label{sold1}
\chi=\pm \left(\frac{1 \pm \tanh(x)}{2}\right)^{1/2}.
\ee
On \eqref{sold1}, $+$ refers to the kink of the right topological sector, and $-$ refers to the left one, as shown on the second graph of Fig.~\ref{plotp6}. The energy density is 
\be
\label{densd1}
\rho(x)=\frac{\sech^2(x)(1\mp \tanh(x))}{\sqrt{64-\beta^2\sech^4(x)(1 \mp \tanh(x))^2}}.
\ee
The potential stability is obtained by means of 
\be
A(x)=\sqrt{\frac{1+\beta^2\phi'^4}{1-\beta^2\phi'^4}}
\ee
and
\be
F_X A=\frac{\sqrt{1+\beta^2\phi'^4}}{1-\beta^2\phi'^4}.
\ee

The quantities presented in this section are depicted in Fig.~\ref{plotp6}. As the parameter $\beta$ increases, the potential maxima tend to become higher and thinner; the energy density increases maintaining the same solutions which are $\beta$-independent; and the stability potential  tends to become an asymmetric volcano. In comparison to what was achieved here, in the kink-like solutions in DBI dynamics \cite{Bazeia:2017mnc}, as the dynamics effects get stronger, the potential maxima tend to become lower and thicker reaching constant values for large  $\chi$, and reducing the energy densities of the solutions. Furthermore, a plateau appeared in the stability potential under DBI dynamics and this does not occur in our arcsin dynamics analyzes, but a volcano behavior arises as provoked by warped geometries \cite{brane,Barbosa}.

Here, for $\beta<<1$, the stability potential develops the following form
\be
U(x) \approx \frac{5}{2} \pm \frac{3}{2}\tanh(x)-\frac{15}{4}\sech(x)^2,
\ee
where it has only one bound state corresponding to $\omega^2 = 0$ \cite{lohe}. This result is equivalent to the one obtained in the $\phi^6$ model in the standard theory.

\subsection{Family 1}

\begin{figure}%
\centering
\includegraphics[scale=0.35]{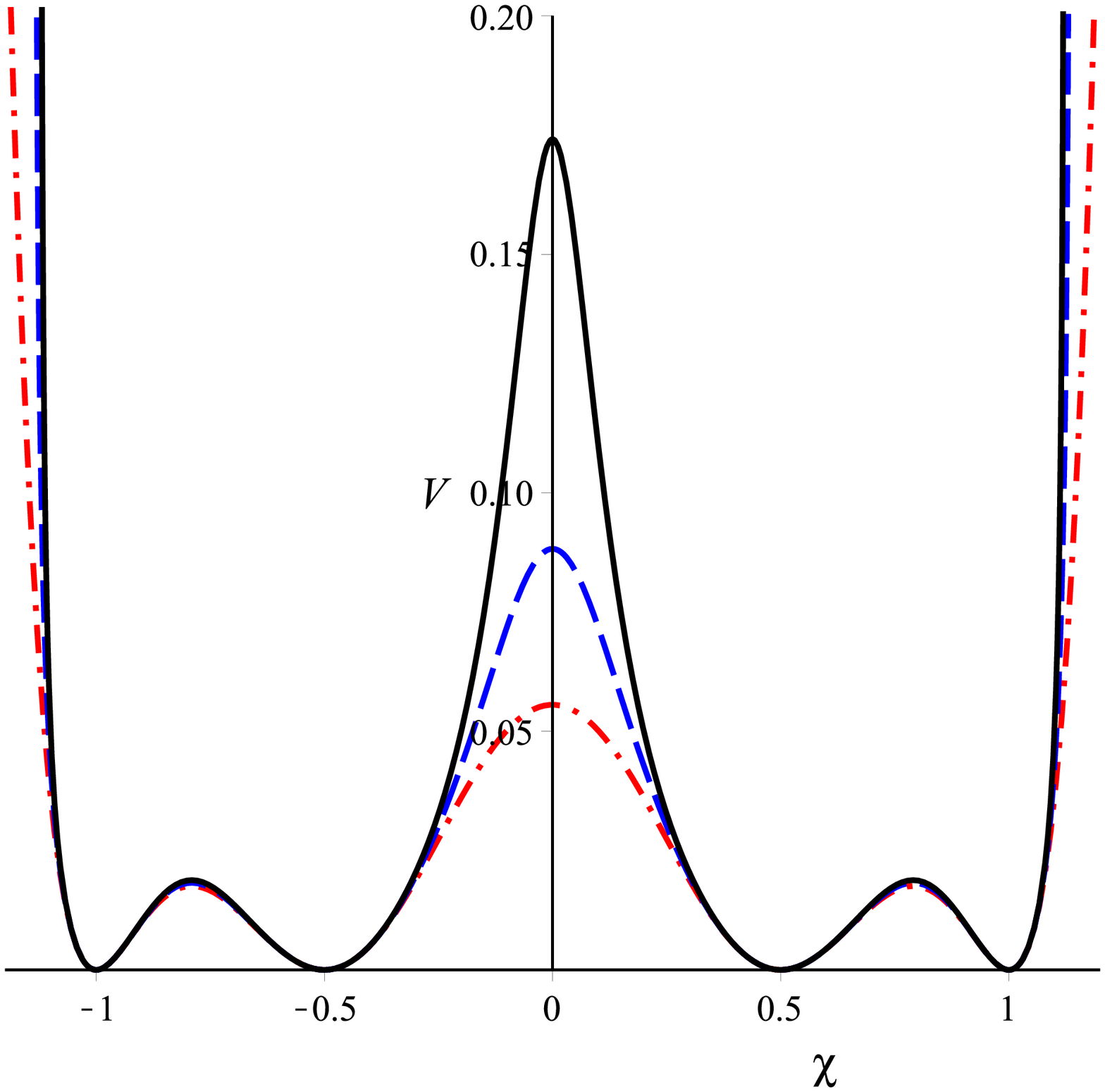}
\includegraphics[scale=0.35]{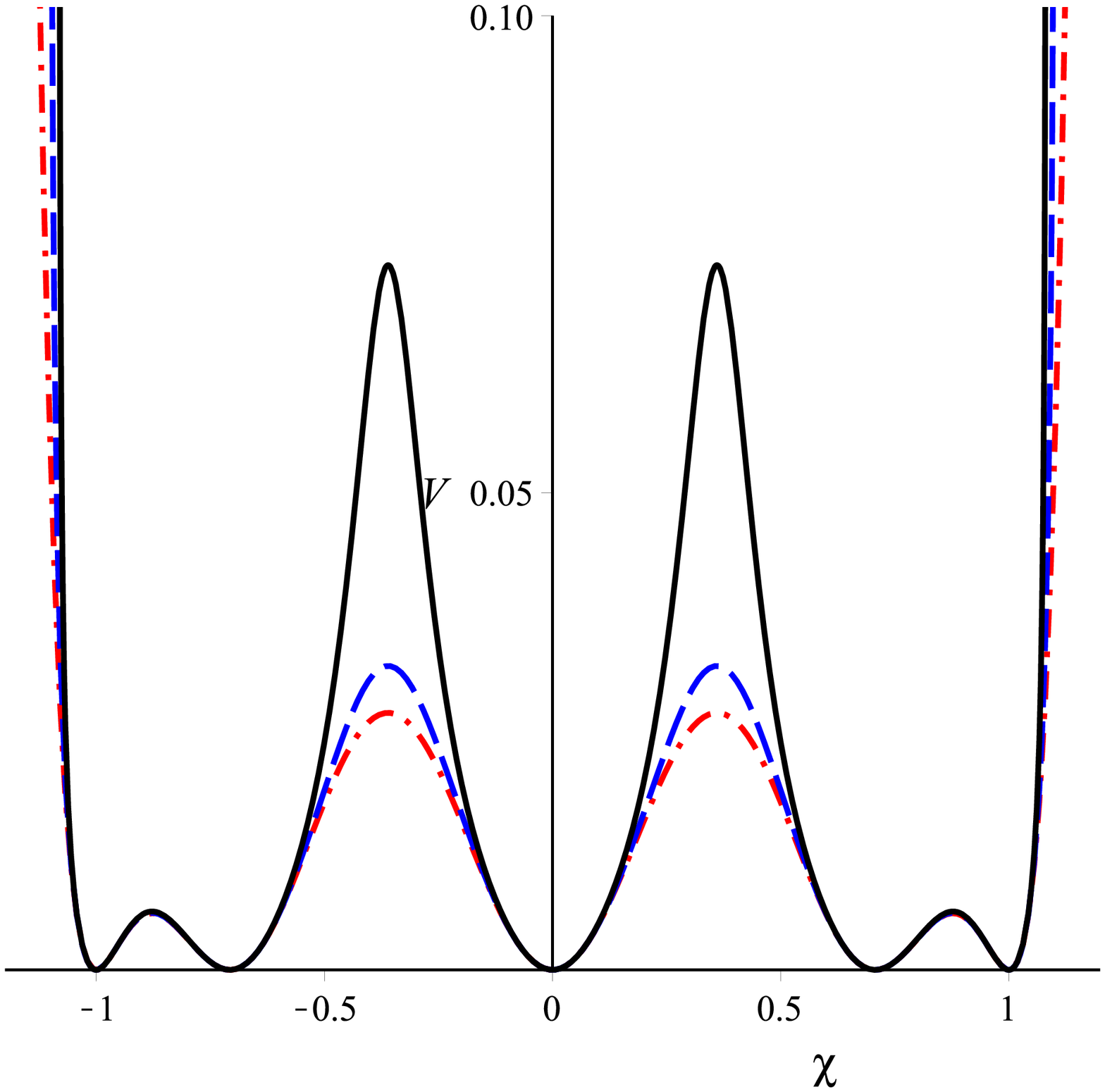}
\caption{The potential \eqref{potdf1}. (a) For $n=3$ and ${\beta} = 0.2, 6, 8$; and (b) for $n=4$ and ${\beta} = 0.2, 8, 16$,  represented by a dash-dotted (red), dashed (blue), and solid (black)  lines, respectively.}
\label{potf1}
\end{figure}

Now, we consider the general deformation function $f(\chi)=\cos(n\arccos(\chi)-k\pi)$,  with $n=1,2,3,4...$ and $k=0,1,2...,2n-1$, which provides a family of potentials including higher order powers of the scalar field \cite{bazeiaLeon,Bazeia:2017mnc}. In this scenario, we have
\be
S_{n}(\chi)=\frac{1}{n}\sqrt{1-\chi^2}\sin(n\,\arccos\chi)\,\cos(k\pi),
\ee
in which in terms of the Chebyshev polynomials becomes 
\be
S_{n}(\chi)=\frac{1}{n}(1-\chi^2)U_{n-1}(\chi)\,\cos(k\pi).
\ee
The deformed potential reads
\ben
\label{potdf1}
\tilde{V_{n}} (\chi)&=&-\frac{1}{2\,\beta}\arcsin{\left(\frac{\beta}{n^2} (1-\chi^2)^2 U_{n-1}^2 (\chi)\right)} \nonumber \\
&&+\frac{(1-\chi^2)^2U_{n-1}^2 (\chi)}{\sqrt{n^4-\beta^2(1-\chi^2)^4 U_{n-1}^4 (\chi)}},
\een
whose minima are $\chi_{min}=\cos\left(\frac{m\pi}{n}\right)$, where $m=0,1,..,n$. The number of minima increases as $n$ increases, being equal to $n+1$, and providing the development of new topological sectors. There are two categories of models: one with a maximum at the origin for $n-$odd and another one with a minimum at the origin for $n-$even. Where the former provides a family of potentials $\tilde{V_{n}}$ that recovers the models \eqref{potp4} and \eqref{potd1} when $n=1,2$, respectively. The potentials for $n=3,4$ are plotted in Fig.~\ref{potf1}. By looking at the standard theory case addressed in the Ref.~\cite{bazeiaLeon},it can beseen that the potentials plotted in the Fig.~\ref{potf1} differ from the standard case expressively as the parameter $\beta$ increases and the field $\chi$ goes near the origin. Considering $\chi=0$, one gets
\ben
\tilde{V_{n}}(0)&=&-\frac{1}{2\,\beta}\arcsin{\left(\frac{\beta}{n^2}  U_{n-1}^2 (0)\right)} +\frac{U_{n-1}^2 (0)}{\sqrt{n^4-\beta^2 U_{n-1}^4 (0)}}. \nonumber \\
\een
Since
\be
U^2_{n-1}(0)=\cos^2\left(\frac{(n-1)\pi}{2}\right)=
\left\{
\begin{array}{c}
0 \mbox{ for $n$ even,} \\ \\
1 \mbox{ for $n$ odd,}
\end{array}
\right.
\ee
Thus, the parameter must be in the range of $0<\beta<n^2$ for $n-$odd, where the origin is a global maximum. 

The solutions are obtained by using the inverse deforming function $f^{-1}(\tanh(x))$, which results in
\be
\chi_{n,k}(x)=\cos\left(\frac{\arccos(\tanh (x))+k\pi}{n}\right).
\label{as}
\ee
Distinct values of $k$ furnish distinct solutions connecting the potential minima, namely $k = 0, . . . , n - 1$ provides the kink solutions, and $k = n, . . . , 2n-1$ provides the anti-kink ones.

The function $S$, in terms of the coordinate $x$, is obtained by applying the solution \eqref{as},
\be
S_{n,k}(x)=\frac{1}{n}\sin\left(\frac{\theta(x)+k\pi}{n}\right)\sech(x),
\ee
where $\theta(x)=\arccos(\tanh (x))$. Then, the energy density \eqref{rhodef} becomes
\be
\rho_{n,k}=\frac{\sin^2\left(\frac{\theta(x)+k\pi}{n}\right)\sech^2(x)}{\sqrt{n^4-\beta^2 \sin^4\left(\frac{\theta(x)+k\pi}{n}\right)\sech^4(x)}}.
\ee
Considering $\beta<<1$, the results presented in this section recovers the ones obtained for a family of polynomial models in the standard model \cite{bazeiaLeon}. 

\subsection{Modified sine-Gordon}

\begin{figure}%
\centering
\includegraphics[scale=0.3]{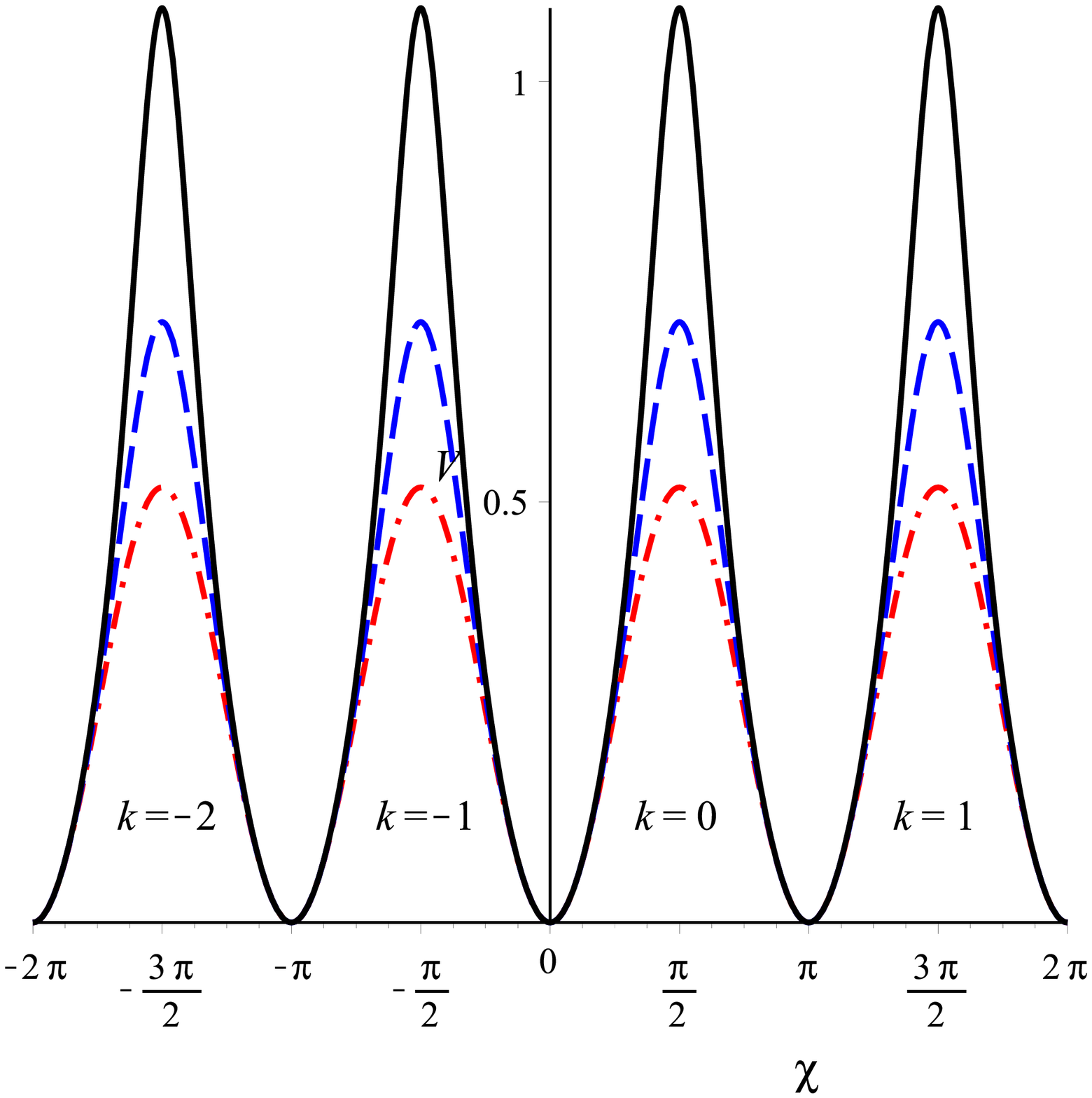}
\includegraphics[scale=0.3]{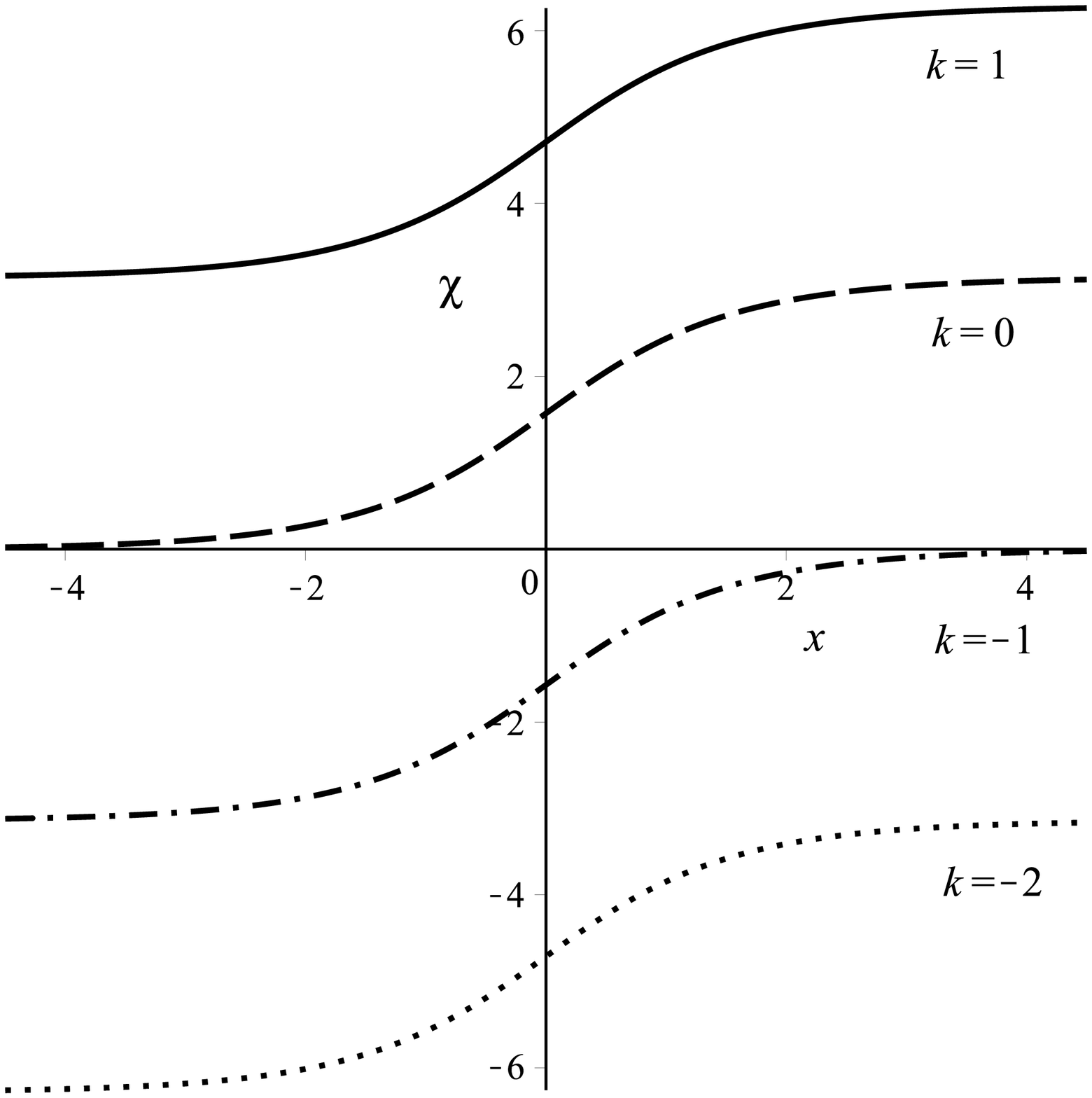}
\includegraphics[scale=0.3]{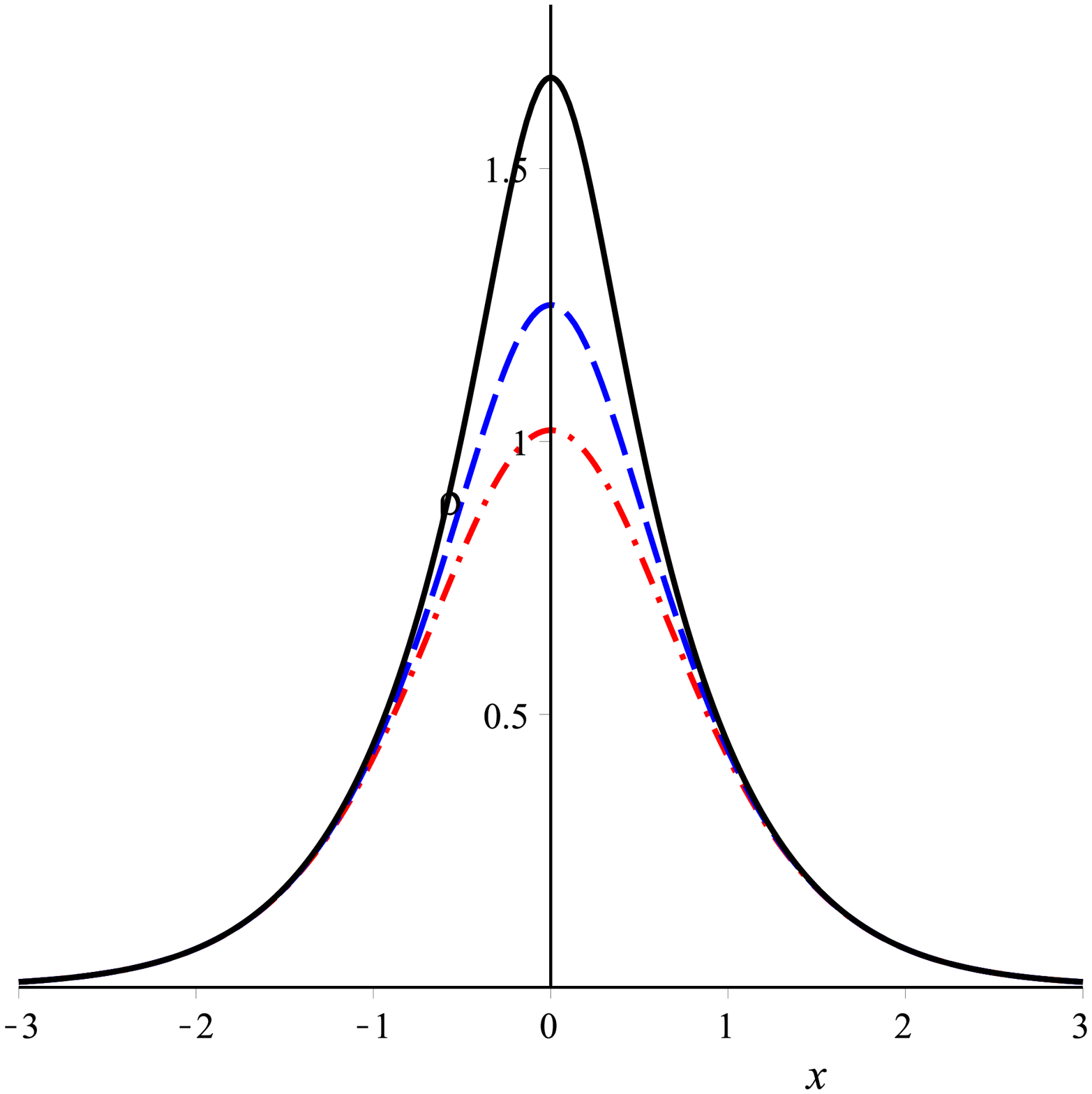}
\includegraphics[scale=0.3]{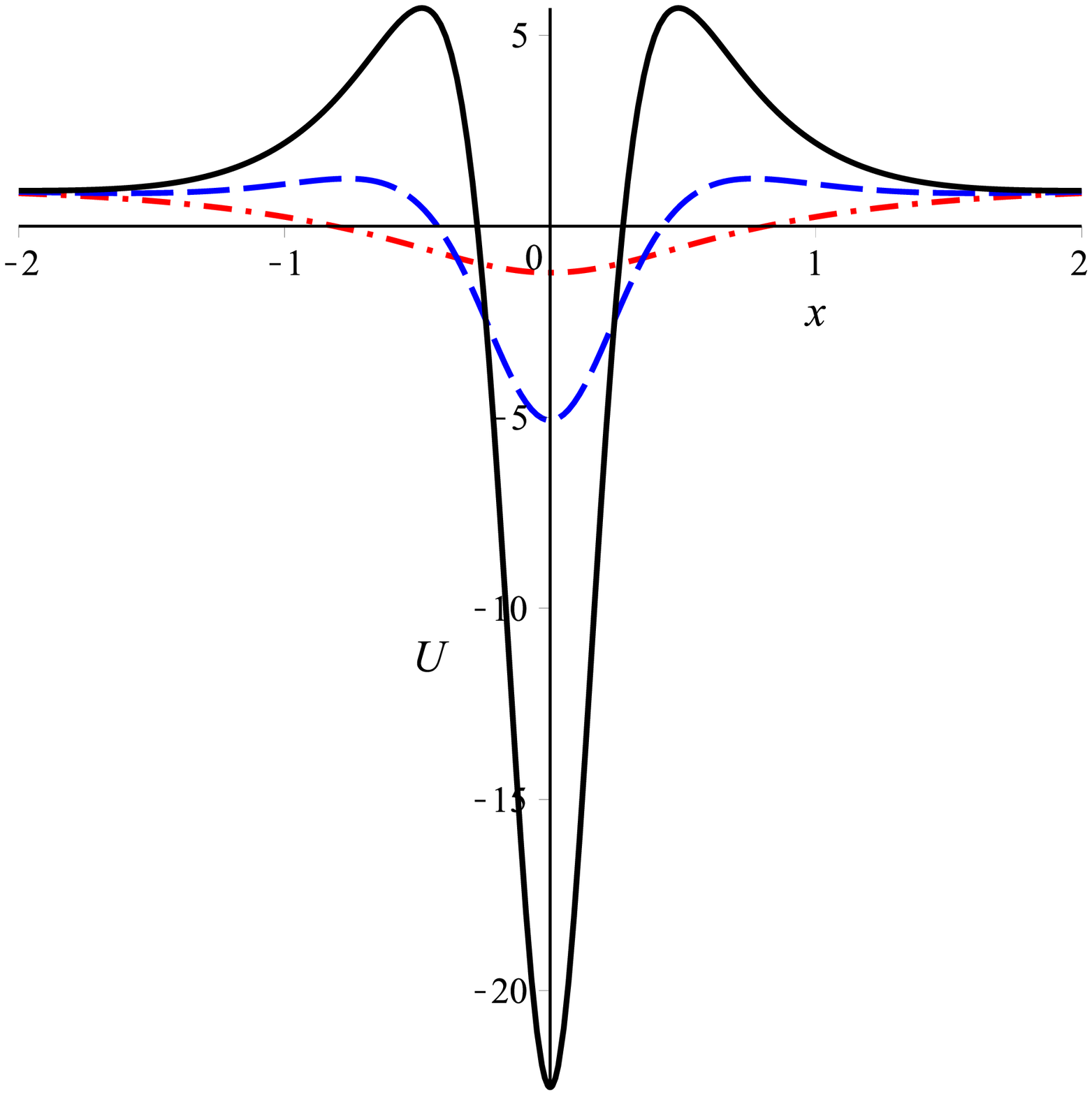}
\caption{The potential \eqref{ptsin}, the kink solutions $\chi_k(x)$ for $k=-2,-1,0,1$, the energy density \eqref{densin}  and the stability potential. For $\beta = 0.2, 0.6, 0.8$, represented by dash-dotted (red), dashed (blue), and solid (black)  lines, respectively.}
\label{plotsin}
\end{figure}

Let us consider the deforming function $f_k(\chi)=-\cos(\chi-k\pi)$, for integer $k$. The function $S(\chi)$, given by Eq.~\eqref{schi}, is
\be
S_k(\chi)=\sin(\chi)\cos(k\pi),
\ee
and the deformed potential becomes
\be
\tilde{V}(\chi)=-\frac{1}{2\beta}\arcsin\left(\beta\sin^2(\chi)\right)+\frac{\sin^2(\chi)}{\sqrt{1-\beta^2\sin^4(\chi)}},
\label{ptsin}
\ee
where $0<\beta<1$. As in the standard sine-Gordon model \cite{tops}, the potential \eqref{ptsin} has identical topological sectors that repeat themselves infinitely, with minima at $\chi_{min}=m\pi$ and maxima at $\chi_{max}=(2m-1)\pi/2$, where $m=0,\pm 1, \pm 2,...$. As the parameter $\beta$ approaches to the unity, the maxima of the potential become sharper, as it can be seen in the first plot of Fig.~\ref{plotsin}. On the other way the solutions that are $\beta-$independent are given by
\be
\chi_k(x)= \left\{
\begin{array}{c}
-\arccos(\tanh(x))+(k+1)\pi \mbox{ for kinks,} \\ \\
\arccos(\tanh(x))+k\pi \mbox{ for anti-kinks.}
\end{array}
\right.
\ee
Each value of $k$ gives a solution connecting distinct sectors, as shown on the second graph of Fig.~\ref{plotsin} for $k=-2,-1,0,1$. The energy density of these solutions is 
\be
\rho(x)=\frac{\sech^2(x)}{\sqrt{1-\beta^2\sech^4(x)}};
\label{densin}
\ee
and the stability potential \eqref{u(z)} is obtained using
\ben
A(x)&=&\sqrt{\frac{1+\beta^2\sech^4(x)}{1-\beta^2\sech^4(x)}}, \\
F_X(x) A(x)&=&\frac{\sqrt{1+\beta^2\sech^4(x)}}{1-\beta^2\sech^4(x)},
\een
as presented in the last plot from Figure~\ref{plotsin}. Here, the stability potential is symmetric and it becomes a volcano as $\beta$ increases. For $\beta<<1$, these results are equivalent to the ones obtained by the usual sine-Gordon model \cite{tops}, and $U(z)$ becomes
\be
U(x) \approx 1-2\,\sech^2(x),
\ee
where the zero mode is only bound state \cite{teller}. 

\subsection{Modified double sine-Gordon} \label{sec-MDSG}

\begin{figure}%
\centering
\includegraphics[scale=0.3]{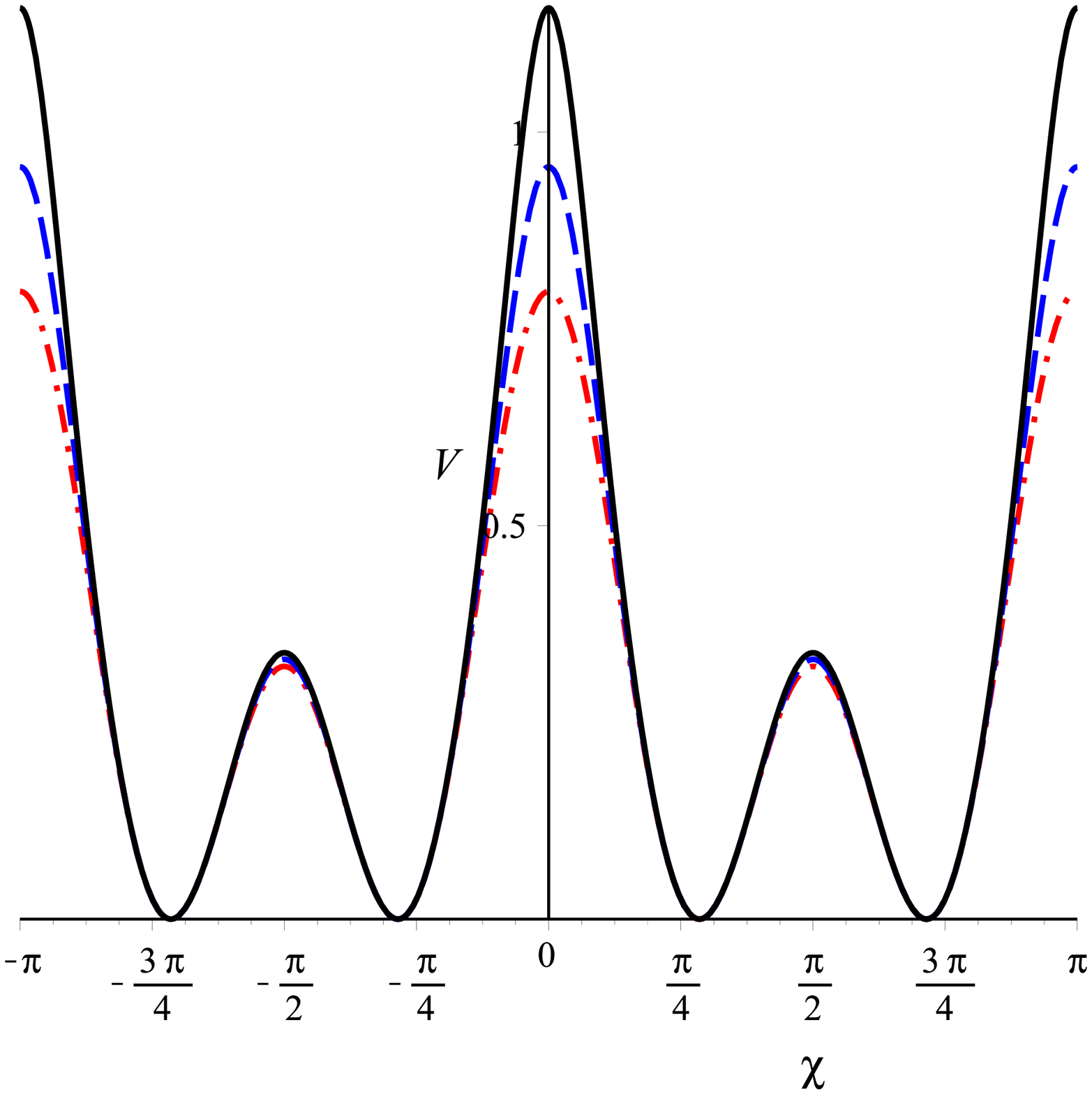}
\includegraphics[scale=0.3]{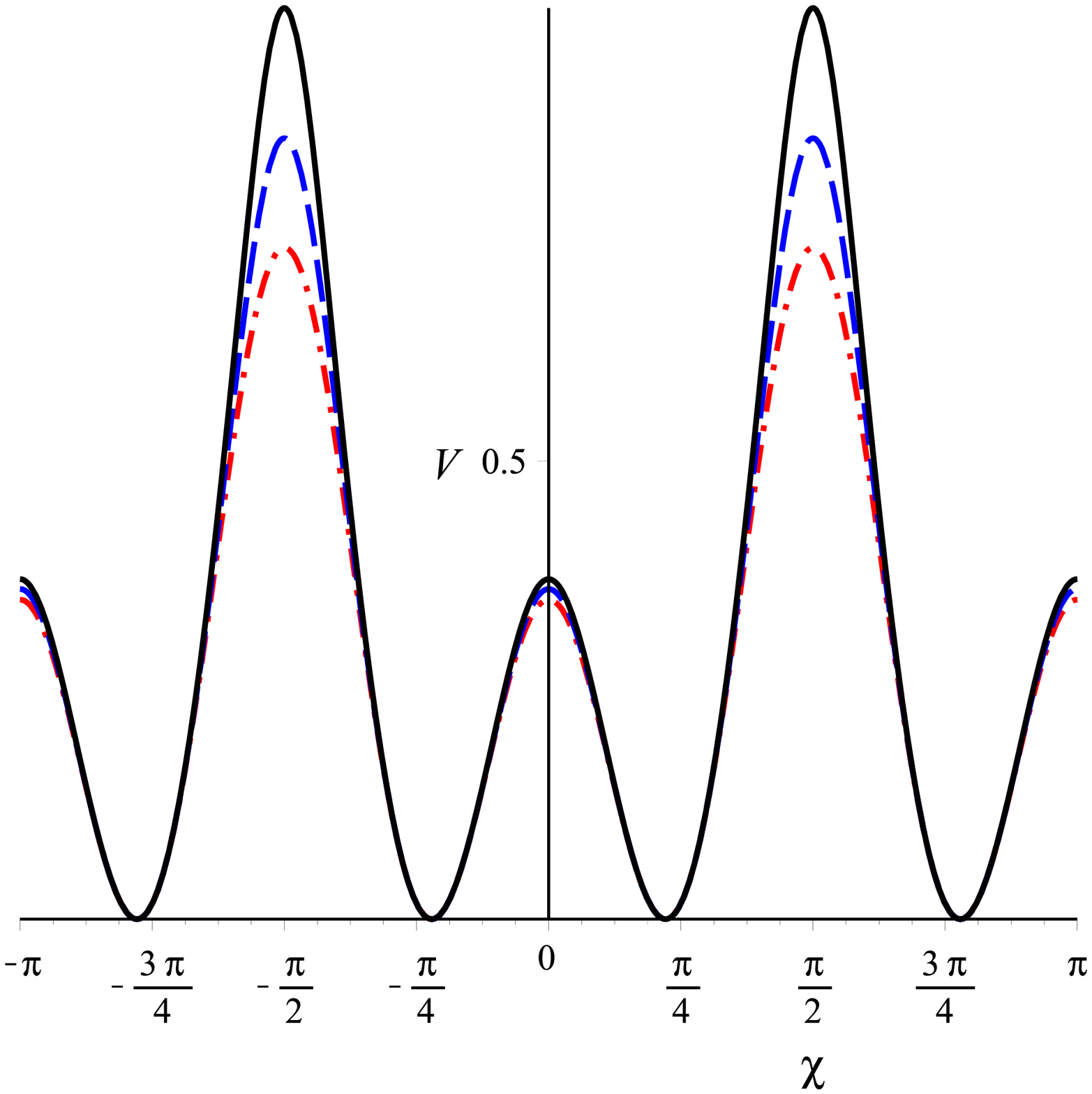}
\caption{The potential \eqref{potdsg}. (a) for $r=0.8$, in the upper panel,  and (b) for $r=1.2$, in the lower panel. Where ${\beta} = 0.1, 0.3, 0.4$ represented by dash-dotted (red), dashed (blue), and solid (black)  lines, respectively.}
\label{plotdoubsin}
\end{figure}
\begin{figure}%
\centering
\includegraphics[scale=0.3]{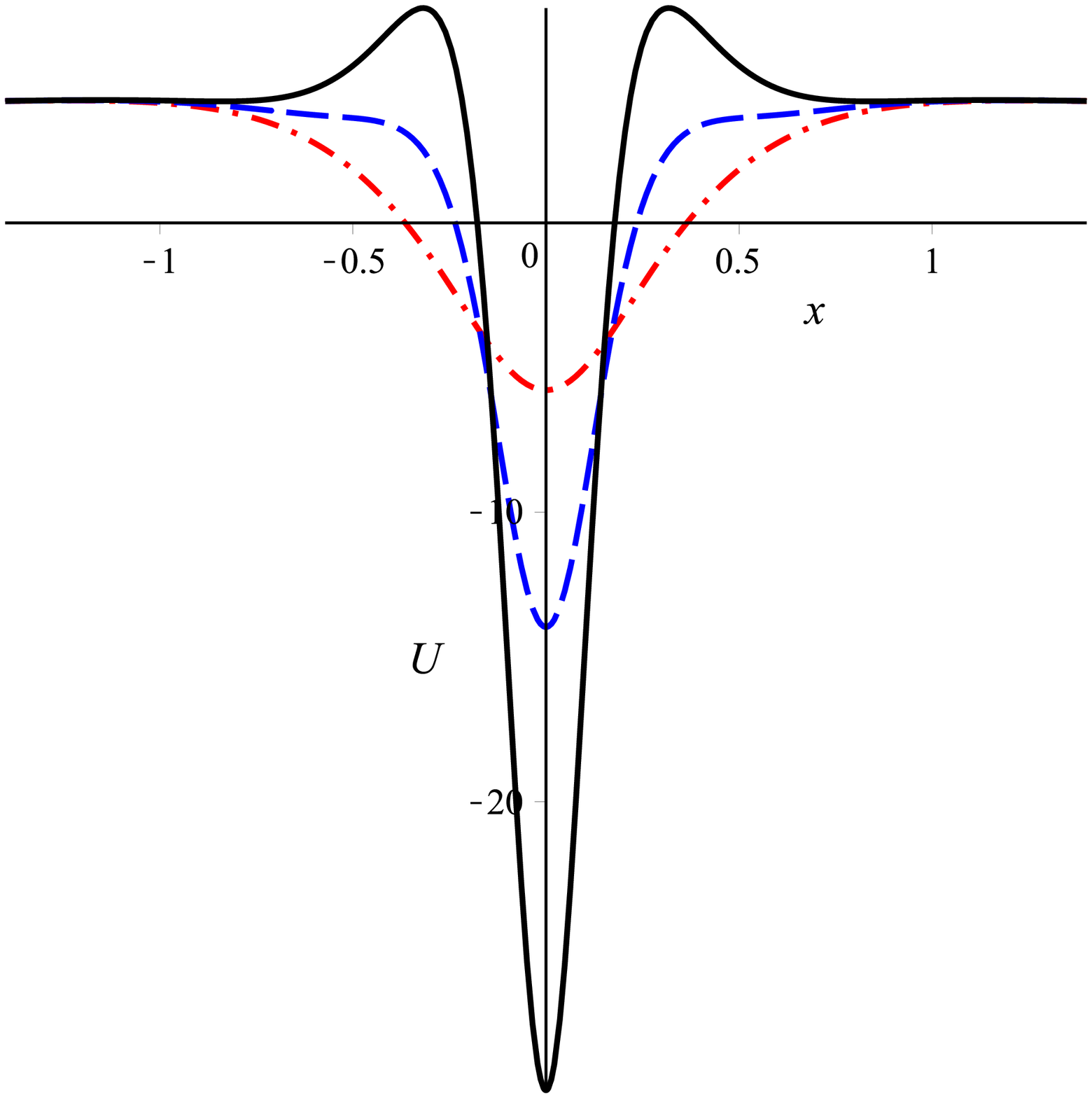}
\includegraphics[scale=0.3]{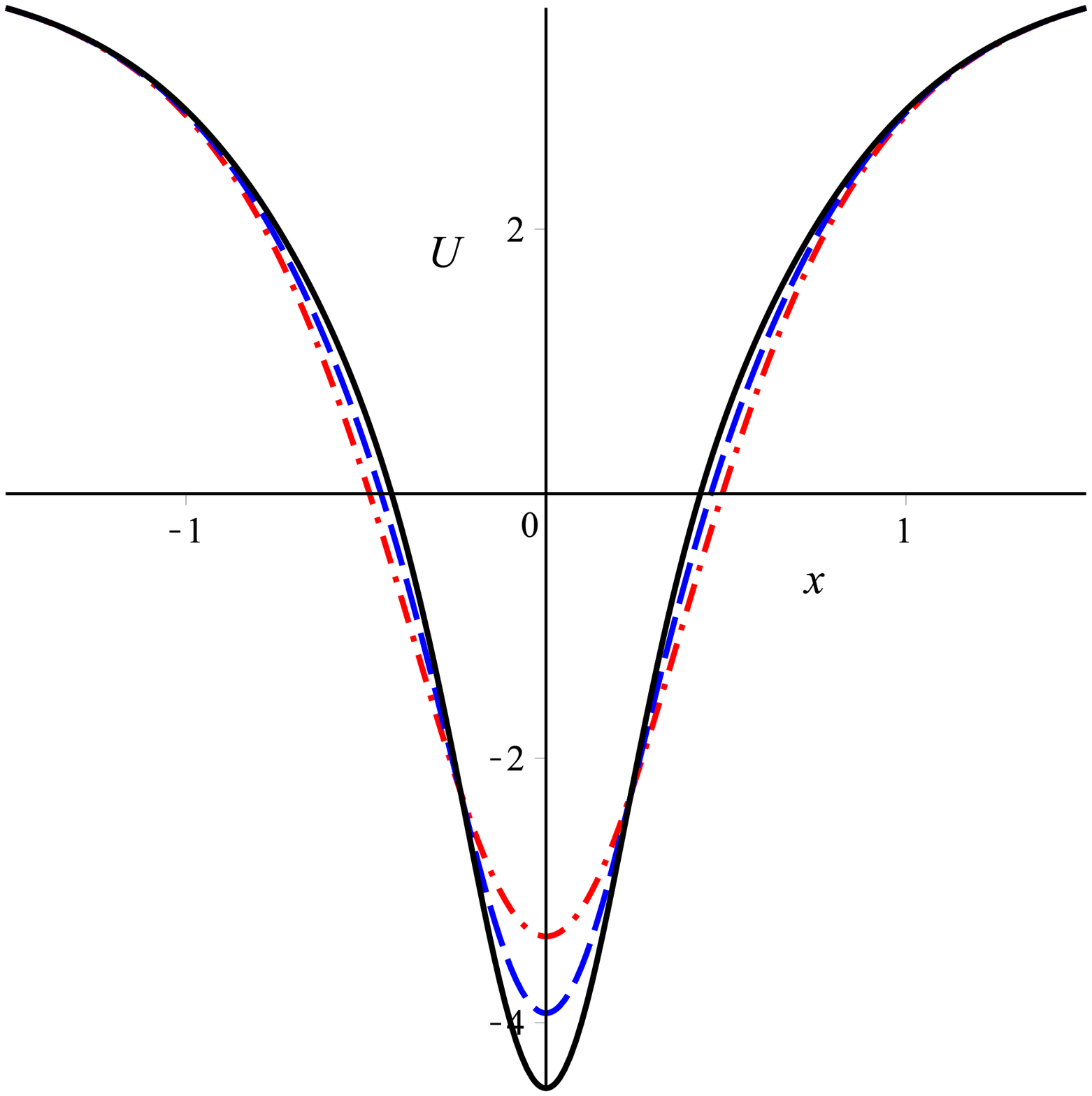}
\caption{The stability potential for the  modified double sine-Gordon model, using $r=0.8$ and the same values of ${\beta}$ used in Fig.~\ref{plotdoubsin}. The upper panel is associated with $\chi_1(x)$, and the lower one with $\chi_2(x)$.}
\label{stabilitydoubsin}
\end{figure}

Note that, if we select the following deforming functions: $f_{r,k}=r\tan(\chi-k\pi)$ and $g_{r,k}=1/r\cot(\chi-k\pi)$, for $r \in (0,\infty)$  and integer $k$ \cite{familySine2009}, then it can be verified 
that the function $S(\chi)$, suggested in Eq.~\eqref{schi}, are the same, which is given by 
\be
S(\chi)=\frac{1}{r}\left[(1+r^2)\cos^2(\chi)-r^2\right].
\ee
The deformed potential \eqref{deformmodel} becomes
\ben
\label{potdsg}
\tilde{V}(\chi)&=&-\frac{1}{2\beta}\arcsin\left(\frac{\beta}{r^2}\left[(1+r^2)\cos^2(\chi)-r^2\right]^2\right) \nonumber\\ &&+\frac{\left[(1+r^2)\cos^2(\chi)-r^2\right]^2}{\sqrt{r^4-\beta^2\left[(1+r^2)\cos^2(\chi)-r^2\right]^4}},
\een
which behaves like a modified double sine-Gordon having two distinct topological sectors for $r\neq 1$ and it acts as a modified sine-Gordon for $r=1$ \cite{doublesine}. The minima are localized at 
\be
\chi_{min}=\pm\arccos\left(\frac{r}{\sqrt{1+r^2}}\right)+m\pi,
\ee
for $m=0,\pm 1, \pm 2, ...$. In addition to that, for $r\in(0,1)$ there are global maxima at $\chi_{max,g}=m\pi$ and local maxima at $\chi_{max,l}=(2m-1)\pi/2$; however this behavior is inverted  for $r>1$, as it can be seen in Fig.~\ref{plotdoubsin}. One more important remark, it is noticed that the non-standard kinetic affects more the global maxima of the potential than the local ones. Particularly,  when $r\in (0,1]$, the origin $\chi=0$ has a global maximum, and the potential becomes
\be
\tilde{V}(0)=-\frac{1}{2\beta}\arcsin\left(\frac{\beta}{r^2}\right)+\frac{1}{\sqrt{r^4-\beta^2}},
\ee
then $\beta$ must be on the interval $0<\beta<r^2$, to obtain  real values of the potential in that point. On the other hand,  when $r>1$, the potential at the global maximum $\chi=\pi/2$ is 
\be
\tilde{V}\left(\frac{\pi}{2}\right)=-\frac{1}{2\beta}\arcsin\left(\beta r^2\right)+\frac{r^2}{\sqrt{1-\beta^2 r^4}},
\ee
and we have $0 < \beta< 1/r^2$. The kink solutions corresponding to the two different sectors of the model are obtained using the inverse deforming functions, $f_{r,k}^{-1}(\phi)$ and $g_{r,k}^{-1}(\phi)$, which gives 
\ben
\chi_1(x)&=&\arctan\left(\frac1r\tanh(x)\right)+k\pi, \\
\chi_2(x)&=&-\arccot(r\tanh(x))+k\pi.
\een
The function $S(\chi)$ evaluated at these solutions becomes
\ben
S_1(x)&=&\frac{r\,\sech^2(x)}{r^2+\tanh^2(x)}, \\
S_2(x)&=&-\frac{r\,\sech^2(x)}{1+r^2\tanh^2(x)};
\een
and the energy densities read
\ben
\rho_1(x)&=&\frac{r^2\sech^4(x)}{\sqrt{(r^2+\tanh^2(x))^4-\beta^2r^4\sech^8(x)}},\\
\rho_2(x)&=&\frac{r^2\sech^4(x)}{\sqrt{(1+r^2\tanh^2(x))^4-\beta^2r^4\sech^8(x)}}.
\een
The stability potentials concerning the two distinct solutions of the model are presented in Fig.~\ref{stabilitydoubsin}, with $r<1$. As it can be seen, as $\beta$ increases, a volcano behavior appears at the larger sector related with $\chi_1(x)$ (first plot). This is the most affected sector by the arcsin dynamics; although, at the smaller one, associated with $\chi_2(x) $, only occurs a slighter variation (second plot).

\subsection{Family 2}

\begin{figure}%
\centering
\includegraphics[scale=0.3]{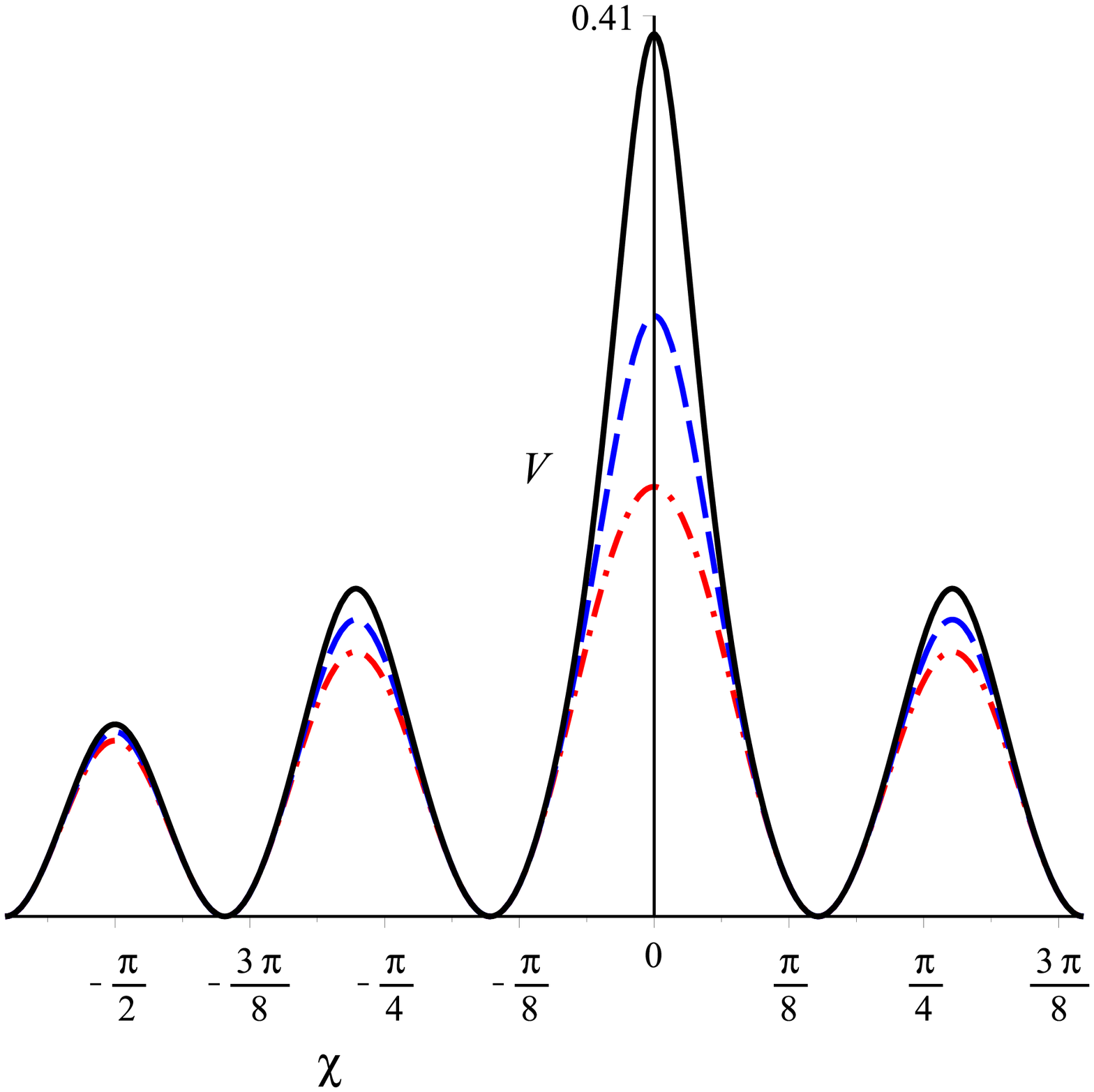}
\includegraphics[scale=0.3]{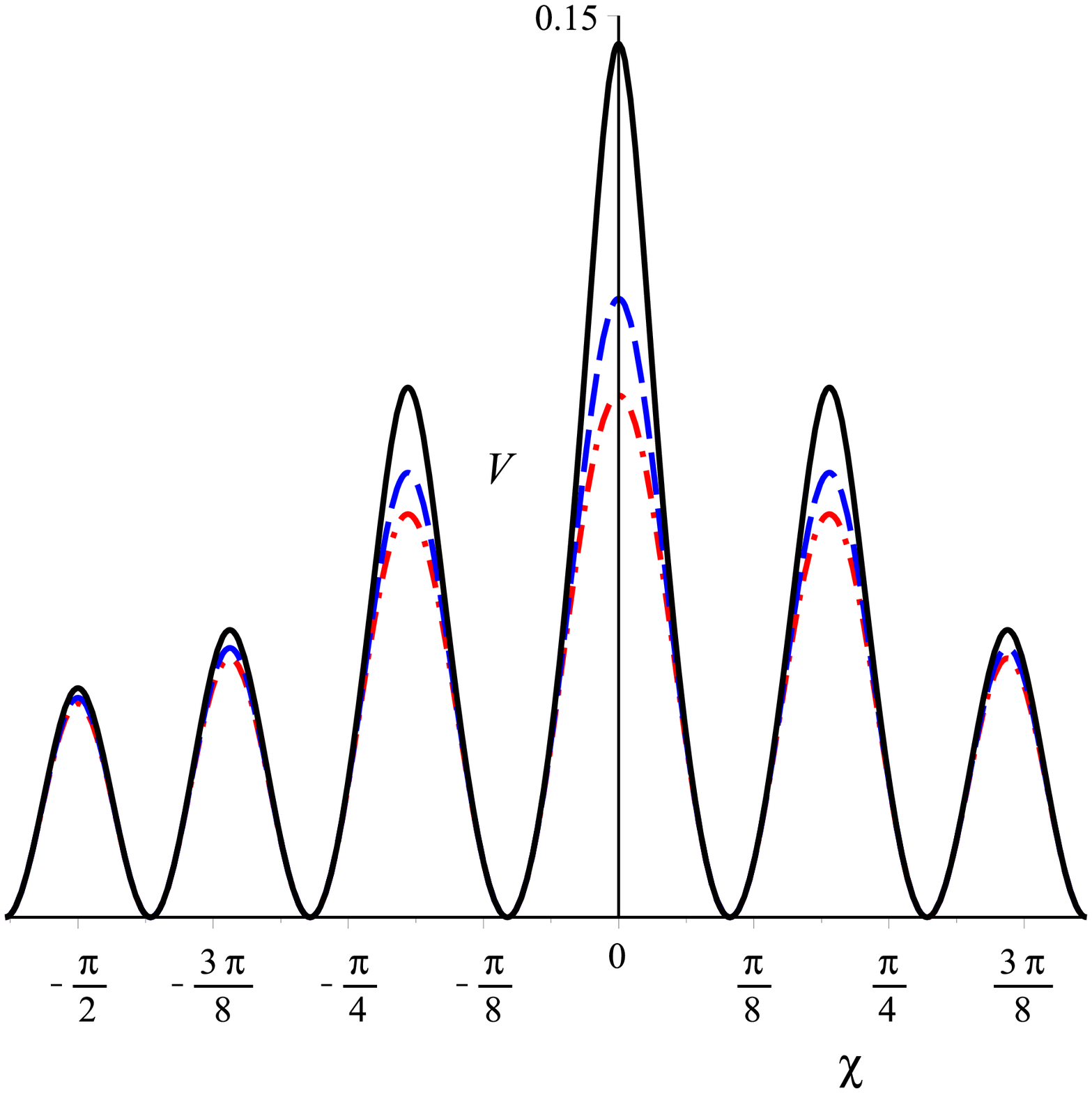}
\caption{The potential \eqref{potfam2} for $r=0.8$, varying $n$ and $\beta$. In the upper panel, 
one uses $n=2$ and $\arctan[(1/r)\tan(3\pi/8)]-\pi \leq \chi \leq \arctan[(1/r)\tan(3\pi/8)]$, with ${\beta} = 0.1, 1.5, 2$, represented by dash-dotted (red), dashed (blue), and solid (black)  lines, respectively. In the lower panel, one uses $n=3$ and $\arctan[(1/r)\tan(5\pi/12)]-\pi \leq \chi \leq \arctan[(1/r)\tan(5\pi/12)]$, with ${\beta} = 0.1, 2.5, 4$, represented by dash-dotted (red), dashed (blue), and solid (black)  lines, respectively.}
\label{plot2family}
\end{figure}

In the last two sections, we have studied models engendering interactions like sine-Gordon. 
Also, there are deforming functions which lead to families of these models, as investigated in Refs.~\cite{familySine2009,Bazeia:2017mnc}. In this sense, we select the functions
\ben
f_{n,r}(\chi)&=&\tan(n\,\arctan(r\tan(\chi-k\pi))-l\pi), \nonumber \\
g_{n,r}(\chi)&=&\cot(n\,\arctan(r\tan(\chi-k\pi))-l\pi), \nonumber \\
h_{n,r}(\chi)&=&\tan\left(n\,\arctan\left(\frac{1}{r}\cot(\chi-k\pi)\right) -l\pi\right), \nonumber\\
q_{n,r}(\chi)&=&\cot\left(n\,\arctan\left(\frac{1}{r}\cot(\chi-k\pi)\right)-l\pi\right). \nonumber \\
\label{functionsfsg}
\een
where  $n=1,2,3,...,$ and $k,l$ are integers, and $r$ is a positive real number. Each one of these functions provides the same potential model, and it furnishes distinct solutions to the novel models. 

The function $S(\chi)$ becomes 
\ben
\label{Snr}
S_{n,r}(\chi)=\pm \frac{1}{nr}\left[(1-r^2) \cos^2(\chi)+r^2\right]\times \nonumber \\ 
\left[2 \,\cos^2(n\arctan(r\tan(\chi)))-1\right].
\een
The deformed potential is given by
\be
\tilde{V}_{n,r}(\chi)=-\frac{1}{2\,\beta}\arcsin{(\beta S_{n,r}^2(\chi))}+\frac{S_{n,r}^2(\chi)}{\sqrt{1-\beta^2 S_{n,r}^4(\chi)}},
\label{potfam2}
\ee
with minima at
\be
\chi_{min}^{n,m,k}=\arctan\left[\frac{1}{r}\tan\left((1+2m)\frac{\pi}{4n}\right)\right]+k\pi,
\ee
where $m=0,...,2n-1$, and $k=0,\pm 1,...,\pm \infty$, as we can see  $\beta$ does not change the minima positions. The parameter $n$ controls the number of distinct topological sectors, which leads to a family of potentials like sine-Gordon. For each value of $n$, there are $2n$ sectors that replicate in a periodic pattern. The replication period of the model is $\pi$, and its periodical structure is localized between the minima $\chi_{min}^{n,n-1,-1}$ and $\chi_{min}^{n,n-1,0}$, where
\be
\chi_{min}^{n,n-1,0}=\arctan\left[\frac{1}{r}\tan\left((2n-1)\frac{\pi}{4n}\right)\right].
\ee
Moreover, among the sectors of these periodic patterns within a $\pi$ interval, $n+1$ are different, as presented in Fig.~\ref{plot2family}, in such a way that $n = 1$ provides a modified double sine-Gordon model with two distinct topological sectors, recovering the previous model explained in Sec.~\ref{sec-MDSG}. The value $n = 2$ originates a modified triple sine-Gordon model, containing three distinct sectors. The quantity $n = 3$ furnishes a modified quadruple sine-Gordon model, having four distinct sectors, and so on. The Figure~\ref{plot2family} depicts the potential \eqref{potfam2} for $n=2,3$ (upper and lower panels, respectively), showing the  arising of new topological sectors as the parameter $n$ increases and how the potential varies as $\beta$ changes. 

At the origin of the potential, $\chi=0$, we have
\be
\tilde{V}(0)=-\frac{1}{2\beta}\arcsin\left(\frac{\beta}{n^2r^2}\right)+\frac{1}{\sqrt{n^4r^4-\beta^2}},
\ee
which is a global maximum, especially when $r<1$. In this case, this restricts $0<\beta<n^2r^2$ so that the model is real at the origin.

In the situation with $r=1$, the potential \eqref{potfam2} becomes a modified sine-Gordon model
{\footnotesize
\ben
\tilde{V}_{n,1}(\chi)&=&-\frac{1}{2\beta}\arcsin\left(\frac{\beta}{n^2}\cos^2(2n\chi)\right)+\frac{\cos^2(2n\chi)}{\sqrt{n^4-\beta^2\cos^4(2n\chi)}}, \nonumber
\een}
with $2n$ similar sectors inside an interval of periodicity $\pi$.

Furthermore, there are four distinct types of solutions related with the inverses of the deforming functions selected in Eq.~\eqref{functionsfsg}. These solutions are
\ben
\label{solFS1}
\chi_{n,l,k}^{(1)}(x)&=&\pm \arctan \left[\frac{1}{r}\tan\left(\frac{\delta(x)+l\pi}{n}\right)\right]+k\pi, \nonumber \\
\label{solFS2}
\chi_{n,l,k}^{(2)}(x)&=&\mp \arctan \left[\frac{1}{r}\tan\left(\frac{\sigma(x)+l\pi}{n}\right)\right]+k\pi, \nonumber \\
\label{solFS3}
\chi_{n,l,k}^{(3)}(x)&=& \mp \arccot \left[r\tan\left(\frac{\delta(x)+l\pi}{n}\right)\right]+k\pi, \nonumber \\
\label{solFS4}
\chi_{n,l,k}^{(4)}(x)&=& \pm \arccot \left[r\tan\left(\frac{\sigma(x)+l\pi}{n}\right)\right]+k\pi, \nonumber \\
\label{solsfam2}
\een
where $\delta(x)=\arctan(\tanh(x))$, and $\sigma(x)=\arccot(\tanh(x))$, then $\delta(x)\in[-\pi/4,\pi/4]$ and $\sigma(x)\in[\pi/4,3\pi/4]$. Let us observe that the solution $\chi_{n,0,0}^{(1)}(x)$ connects the minima of the central topological sector, belonging the interval $\left[ -\arctan \left((1/r)\tan\left(\pi/4n \right)\right), \arctan \left((1/r)\tan\left(\pi/4n \right)\right)\right]$, \textit{i.e.}, between $\chi_{min}^{n,2n-1,0}$ and $\chi_{min}^{n,0,0}$.

Besides that, selecting the parameters $n$ and $r$, each value of the indexes $l$ and $k$ furnishes solutions among the several sectors of $\tilde{V}_{n,r}(\chi)$. The  corresponding energy densities are given by
\be
\rho_{n,r,l,k}^{(i)}(x)=\frac{\left(S_{n,r,l,k}^{(i)}(x)\right)^2}{\sqrt{1-\beta^2\left(S_{n,r,l,k}^{(i)}(x)\right)^4}},
\ee
where $i=1,2,3,4$, and the functions $S_{b,r,l,k}^{(i)}(x)$ are evaluated in the four distinct kinds of solutions described in the expressions \eqref{solsfam2}, that are
\ben
\label{SFS}
S_{n,r,l,k}^{(1)}(x)&=&\frac{\sech(x)^2\sec\left(\frac{\delta+l\pi}{n}\right)^2}{n\,r\left(1+\tanh(x)^2\right)\left(1+\dfrac{1}{r^2}\tan\left(\frac{\delta+l\pi}{n}\right)^2\right)}, \nonumber \\
S_{n,r,l,k}^{(2)}(x)&=&-\frac{\sech(x)^2\sec\left(\frac{\sigma+l\pi}{n}\right)^2}{n\,r\left(1+\tanh(x)^2\right)\left(1+\dfrac{1}{r^2}\tan\left(\frac{\sigma+l\pi}{n}\right)^2\right)}, \nonumber \\
S_{n,r,l,k}^{(3)}(x)&=&-\frac{r\,\sech(x)^2\sec\left(\frac{\delta+l\pi}{n}\right)^2}{n\left(1+\tanh(x)^2\right)\left(1+r^2\tan\left(\frac{\delta+l\pi}{n}\right)^2\right)}, \nonumber\\
S_{n,r,l,k}^{(4)}(x)&=&\frac{r\,\sech(x)^2\sec\left(\frac{\sigma+l\pi}{n}\right)^2}{n\left(1+\tanh(x)^2\right)\left(1+r^2\tan\left(\frac{\sigma+l\pi}{n}\right)^2\right)}. \nonumber \\
\een
If we assume the limit $\beta \rightarrow 0$, we recover the results obtained in Ref.~\cite{Bazeia:2017mnc}. In this case, the energy densities of the solutions will be given through the functions $S_{b,r,l,k}^{(i)}(x)$.

\section{Comments and Conclusions}\label{sec-com}

In this work, we have investigated the presence of kink structures in two-dimensional scalar field models, with a non-linear kinetic term inspired by the work \cite{Kruglov:2014iwa}.
In order to achieve our results, we have used the first-order formalism and the deformation procedure \cite{FOFGD,Bazeia:2002xg,BLM2014}, which were useful in this paper to study analytically the presence of kink structures in a real scalar model inspired by the non-linear electrodynamics proposed in Ref.~\cite{Kruglov:2014iwa}. One of the most significant advantages of the first-order framework is that it allows us to obtain topological solutions by reducing the second-order equation of motion into first-order differential equations \cite{FOFGD}. Furthermore, the deformation method furnishes novel analytical models, given an initial model of known solutions \cite{Bazeia:2002xg,BLM2014}. 

We have analyzed two kinds of scalar field models, one described by polynomial-like interactions and other by non-polynomial ones. Both models were modified by the nonstandard dynamics. We show that, the non-linearities of the models are incorporated in a way that the arcsin theory preserves the kink-like behavior, although it changes quantities such as energy density and stability potential. Furthermore, these non-linearities are controlled by the parameter $\beta$, and the topological solutions do not depend on $\beta$, as it occurs in standard and DBI models \cite{Bazeia:2017mnc}; but from the equations \eqref{rhoarcsin} and \eqref{rhodef}, and from the Figures~\ref{plot} to \ref{stabilitydoubsin}, we can see that both energy density and the potential depends on $\beta$.  

As it was pointed out by the expansion \eqref{expansion}, we recover the canonical theory for small values of $\beta$. The plots presented in the Fig.~\ref{plot} reveal that the classical potential, the energy density, and the stability potential diverges, as ${\beta}$ approaches the unity. As can be seen in the  Figures~\ref{plot} to \ref{stabilitydoubsin},  the arcsin dynamics originates a volcano behavior on the stability potential,  such feature also can
also be produced by warped geometries \cite{brane,Barbosa}.

Therefore, we have shown that analytical solutions for kink structures, driven by arcsin scalar dynamics, can be found. This is shown in a such a way that, the kinetics term and the classical potential are modified. As a result we have that the kink solutions are preserved and the energy densities and stability potentials vary according to the dimensional parameter inherited to non-linear model. This perspective is one first step for more general scenarios with different dynamics and different interactions, using the deformation procedure and the first-order formalism \cite{FOFGD,Bazeia:2002xg,BLM2014}.

Furthermore, although this work deals with the description of kinklike structures in a generalized scalar field theory, it can also be related to several other situations of physical interest. In \cite{Bazeia:2017nas}, the authors show how the first-order formalism can be used to describe gauge systems with vortices. Modified Lagrangian densities are used to investigate generalizations of Maxwell-Higgs, Chern-Simons-Higgs, and Maxwell-Chern-Simons-Higgs vortices, enabling then solutions based on the first-order framework. In a forthcoming paper, we show how the theoretical framework developed here, plus what was developed in \cite{Bazeia:2017nas}, can be used to present how the first-order formalism can be implemented in more general electromagnetic scenarios, including, for instance, high-order derivatives gauge theories. Besides this, the setup presented here is also being used to investigate the application of the first-order framework in describing the holographic superconductor presented in \cite{Kruglov:2018jee}.

For more future applications, we have that, by following the routes presented in \cite{braneworldGD, BazeiaBranes}, extensions for scalar field systems in braneworlds scenarios can be found. Extensions for complex scalar fields can also be achieved. In \cite{Bazeia:2017irs}, the complete factorization of the equations of motion for a complex scalar field system engendered by non-canonical dynamics is worked out (which can be seeing as a general scalar theory composed of two real scalar fields). The generalization of the first-order formalism to multi-component scalar fields is presented in \cite{Granado:2019bky}, and as it is pointed out there, applications to cosmology can also be found. In \cite{Ignatev:2017vww}, the authors present a non-linear multi-component scalar cosmological model, and the setup presented here can be investigated in this model. 

\section*{Acknowledgments}
The authors thank D. Bazeia and R. Menezes for the early discussions. 


\begin{thebibliography}{99}

\bibitem{Born:1934gh} 
  M.~Born and L.~Infeld,
  Proc.\ Roy.\ Soc.\ Lond.\ A {\bf 144}, no. 852, 425 (1934).

\bibitem{Dirac:1962iy}
  P.~A.~M.~Dirac,
  Proc.\ Roy.\ Soc.\ Lond.\ A {\bf 268} (1962) 57.

\bibitem{Kruglov:2007bh}
  S.~I.~Kruglov,
  Phys.\ Rev.\ D {\bf 75}, 117301 (2007).

\bibitem{Kruglov:2001dp}
  S.~I.~Kruglov,
  Annals Phys.\  {\bf 293}, 228 (2001)
  [hep-th/0110061].

\bibitem{Kruglov:2007zr}
  S.~I.~Kruglov,
  Phys.\ Lett.\ B {\bf 652}, 146 (2007)
  [arXiv:0705.0133 [hep-ph]].

\bibitem{Gaete:2013dta}
  P.~Gaete and J.~Helayel-Neto,
  Eur.\ Phys.\ J.\ C {\bf 74}, no. 3, 2816 (2014)
  [arXiv:1312.5157 [hep-th]].

\bibitem{Kruglov:2014hpa} 
  S.~I.~Kruglov,
  Annals Phys.\  {\bf 353}, 299 (2014)
  [arXiv:1410.0351 [physics.gen-ph]]

\bibitem{Kruglov:2014iwa} 
  S.~I.~Kruglov,
  Annalen Phys.\  {\bf 527}, 397 (2015)
  [arXiv:1410.7633 [physics.gen-ph]].

\bb{tops}
R. Rajaraman, {\it{Solitons and Instantons}} (North-Holland, 1982); A. Vilenkin and E. P. S. Shellard, {\it{Cosmic Strings and
Others Topological Defects}} (Cambridge, 1994); N. Manton and P. Sutcliffe, {\it{Topological Solitons}} (Cambridge, 2004); P. Dorey, K. Mersh, T. Romanczukiewicz, and Ya. Shnir, Phys. Rev. Lett. 107, 091602 (2011); M. Kawasaki, K. Saikawa,
and T. Sekiguchi, Phys. Rev. D 91, 065014 (2015); R. P. L. Azevedo and C. J. A. P. Martins, Phys. Rev. D 95, 043537
(2017).P. M. Chaikin and T. C. Lubensky, {\it{Principles of Condensed Matter Physics}} (Cambridge University Press, Cambridge, England,
1995); D. Walgraef, {\it{Spatio- Temporal Pattern Formation}} (Springer-Berlag, New York, 1997); G. Bertotti, Hysteresis
in Magnetism (Academic, San Diego, CA, 1998).
A. Vanhaverbeke, A. Bischof, and R. Allenspach, Phys. Rev. Lett. 101, 107202 (2008); J. C. Y. Teo and C. L. Kane,
Phys. Rev. B 82, 115120 (2010); D. Claudio-Gonzalez, A. Thiaville, and J. Miltat, Phys. Rev. Lett. 108, 227208 (2012);
F.J. Buijnsters, A. Fasolino, and M.I. Katsnelson, Phys. Rev. Lett. 113, 217202 (2014); A. Bauer at al. Phys. Rev. B 95,
024429 (2017); A. O. Sorokin, Phys. Rev. B 95, 094408 (2017).


\bb{bogomol}
E. B. Bogomol'Nyi, 
Sov. J. Nucl. Phys {\bf 24}, 449 (1976); 
M. K. Prasad, C. M. Sommerfield, 
Phys. Rev. Lett. {\bf 35}, 760 (1975).

\bb{FOFGD}
D. Bazeia, L. Losano and R. Menezes, Phys. Lett. B {\bf668}, 246 (2008).   D.~Bazeia, L.~Losano, R.~Menezes and J.~C.~R.~E.~Oliveira,   Eur.\ Phys.\ J.\ C {\bf 51}, 953 (2007).


\bibitem{FIRST} 
  D.~Bazeia, L.~Losano, M.~A.~Marques, R.~Menezes and I.~Zafalan,
  arXiv:1708.07754 [hep-th];
  D.~Bazeia, L.~Losano and J.~J.~Rodrigues,
  Int.\ J.\ Theor.\ Phys.\  {\bf 54}, no. 6, 2087 (2015); 
   D.~Bazeia, A.~S.~Lobao, L.~Losano and R.~Menezes,
  Eur.\ Phys.\ J.\ C {\bf 74}, no. 2, 2755 (2014);
D.~Bazeia, A.~S.~Lobao, L.~Losano and R.~Menezes,
  Phys.\ Rev.\ D {\bf 88} (2013) 045001;
  D.~Bazeia, L.~Losano, J.~J.~Rodrigues and R.~Rosenfeld,
  Eur.\ Phys.\ J.\ C {\bf 55}, 113 (2008);
  D.~Bazeia, L.~Losano and J.~J.~Rodrigues,
  hep-th/0610028;
   V.~I.~Afonso, D.~Bazeia and L.~Losano,
  Phys.\ Lett.\ B {\bf 634}, 526 (2006);
   D.~Bazeia, C.~B.~Gomes, L.~Losano and R.~Menezes,
  Phys.\ Lett.\ B {\bf 633}, 415 (2006).

 

\bibitem{Bazeia:2017mnc} 
  D.~Bazeia, E.~E.~M.~Lima and L.~Losano,
  Annals Phys.\  {\bf 388}, 408 (2018)

\bibitem{Bazeia:2002xg} 
  D.~Bazeia, L.~Losano and J.~M.~C.~Malbouisson,
  Phys.\ Rev.\ D {\bf 66}, 101701 (2002)

\bb{BLM2014}
D. Bazeia, L. Losano and R. Menezes, Phys. Lett. B {\bf 731}, 293 (2014).


\bb{braneworldGD}
D. Bazeia, A. R. Gomes, L. Losano, and R. Menezes, Physics Letters B, {\bf 671}, 402 (2009). 

\bb{BazeiaBranes}
D. Bazeia, F. A. Brito, and J. R. Nascimento, Phys. Rev. D {\bf 68}, 085007 (2003); 
V.I. Afonso, D. Bazeia, and L. Losano, Phys. Lett. B {\bf634}, 526 (2006);
D. Bazeia, A. S. Lobao Jr., and R. Menezes, Phys. Lett. B {\bf 743}, 98 (2015).


\bibitem{Granado:2019bky}
  D. R. Granado,
 EPL {\bf 126} (2019) no.6,  61001

\bibitem{Bazeia:2017irs} 
  D.~Bazeia, D.~R.~Granado and E.~E.~M.~Lima,
EPL {\bf 119}, no. 6, 61002 (2017)
  
\bibitem{Gies:2006ca} 
  H.~Gies, J.~Jaeckel and A.~Ringwald,
  Phys.\ Rev.\ Lett.\  {\bf 97}, 140402 (2006)
  [hep-ph/0607118].

\bibitem{teller}	
N. Rosen and P. M. Morse, Phys. Rev. {\bf 42}, 210 (1932); G. Poschl and E. Teller, Z. Phys. {\bf 83}, 143 (1933).

\bb{brane}
D. Bazeia, C. Furtado, A. R. Gomes,
Journal of Cosmology and Astroparticle Physics {\bf 2004}, 002 (2004);
D. Bazeia, A. R. Gomes,
Journal of High Energy Physics {\bf 2004}, 012 (2004).

\bb{Barbosa}
Barbosa-Cendejas, Nandinii, and Alfredo Herrera-Aguilar,
 Physical Review D {\bf 73}, 084022 (2006).


\bb{lohe}
M. A. Lohe, 
Physical Review D {\bf 20}, 3120 (1979).

\bb{bazeiaLeon}
D. Bazeia, M. G. Leon, L. Losano, and J. M. Guilarte,
Physical Review D {\bf 73}, 105008 (2006).

\bb{familySine2009}
D. Bazeia, L. Losano, R. Menezes, M.A.M. Souza, Europhys. Lett. 87 (2009) 21001.

\bb{doublesine}
C.A. Condat, R.A. Guyer, M.D. Miller, Phys. Rev. B 27 (1983) 474;
D.K. Campbell, M. Peyrard, P. Sodano, Physica D 19 (1986) 165;
G. Mussardo, V. Riva, G. Sotkov, Nuclear Phys. B 687 (2004) 189;
D. Bazeia, L. Losano, R. Menezes, Physica D 208 (2005) 236;
G. Takacs, F. Wagner, Nuclear Phys. B 741 (2006) 353.

\bibitem{Bazeia:2017nas} 
  D.~Bazeia, L.~Losano, M.~A.~Marques, R.~Menezes and I.~Zafalan,
  Nucl.\ Phys.\ B {\bf 934}, ('212 (2018)

\bibitem{Kruglov:2018jee} 
  S.~I.~Kruglov,
  Annalen Phys.\  {\bf 530}, no. 8, 1800070 (2018)

\bibitem{Ignatev:2017vww} 
  Y.~G.~Ignat'ev and A.~A.~Agathonov,
Grav.\ Cosmol.\  {\bf 23}, no. 3, 230 (2017)

\end{thebibliography}
\end{document}